\documentclass[10pt]{iopart}

\newcommand{\ket}[1]{\left\vert#1\right\rangle}

\newcommand{\bra}[1]{\left\langle#1\right\vert}

\newcommand{\set}[1]{\left\{#1\right\}}
\usepackage{graphicx,caption}
\usepackage{xcolor}

\usepackage{iopams}
\expandafter\let\csname equation*\endcsname\relax

\expandafter\let\csname endequation*\endcsname\relax

\usepackage{amsmath}

\renewcommand{\text}[1]{\mathrm{#1}}
\usepackage[colorlinks,urlcolor=blue,linkcolor=blue,citecolor=blue]{hyperref}

\usepackage{dsfont}

\usepackage{footmisc}
\usepackage{mathrsfs}

\usepackage{calc}
\usepackage{comment}
\usepackage[normalem]{ulem}

\newcommand{\beq}{\begin{equation}}
\newcommand{\eeq}{\end{equation}}

\usepackage{cite}
\usepackage{pifont}
\usepackage{bbm}
\usepackage{comment}
\usepackage{graphicx}
\usepackage{bm}

\graphicspath{{Figures/}}

\begin{document}
\title{Adaptive Bayesian phase estimation for quantum error correcting codes}
\author{F. Mart\' inez-Garc\'ia, D. Vodola, and M. M\"uller}
\address{Department of Physics, Swansea University, Singleton Park, Swansea SA2 8PP, United Kingdom.}

\begin{abstract}
	
	Realisation of experiments even on small and medium-scale quantum computers requires an optimisation of several parameters to achieve high-fidelity operations. As the size of the quantum register increases, the characterisation of quantum states becomes more difficult since the number of parameters to be measured grows as well and finding efficient observables in order to estimate the parameters of the model becomes a crucial task. Here we propose a method relying on application of Bayesian inference that can be used to determine systematic, unknown phase shifts of multi-qubit states. This method offers important advantages as compared to Ramsey-type protocols. First, application of Bayesian inference allows the selection of an adaptive basis for the measurements which yields the optimal amount of information about the phase shifts of the state. Secondly, this method can process the outcomes of different observables at the same time. This leads to a substantial decrease in the resources needed for the estimation of phases, speeding up the state characterisation and optimisation in experimental implementations. The proposed Bayesian inference method can be applied in various physical platforms that are currently used as quantum processors.
\end{abstract}

\maketitle

\section{Introduction}

Quantum computers have the potential to solve some computationally hard problems in a more efficient way than classical computers \cite{nielsen2002quantum}. However, due to coupling with the environment, they are more susceptible than their classical counterparts to dynamical errors that affect the correct behaviour of the algorithms performed \cite{divincenzo2000physical}. In order to cope with dynamical errors, quantum error correction techniques~\cite{terhal2015quantum} need to be applied together with a correct initialisation of quantum states that, in general, suffers from different types of noise. These imperfections can often be modelled  as irreversible couplings to the environment~\cite{breuer2002theory} or as unknown but constant unitary operations appearing due to systematic errors. Due to their constant nature, the latter can be compensated by determining the unknown operations and applying their inverse onto the state. The simplest instance of such systematic errors is given by single-qubit phase shifts which can transform a desired state $\alpha \ket{0} + \beta \ket{1}$ into $\alpha \ket{0} + \beta e^{i\phi} \ket{1}$ where $\phi$ is an unknown but constant phase. Estimates of this phase shift can be obtained by performing Ramsey-type experiments\cite{ramsey1950molecular,foot2005atomic} and, more recently, by adaptive methods based on application of Bayesian inference~\cite{huszar2012adaptive,wiebe2016efficient,granade2017practical,sugiyama2012adaptive,fischer2000quantum,kravtsov2013experimental,mahler2013adaptive,granade2012robust}. These adaptive methods select the measurement to be performed by numerical optimisation of the information gained based on the results obtained in the previous measurements.

The characterisation of multi-qubit states, such as those needed for quantum error correcting codes \cite{lidar2013quantum}, is a more complex problem. It is well known that quantum state tomography \cite{cramer2010efficient} becomes impractical to fully characterise these states since the resources needed scale exponentially with the number of qubits. Additionally, the systematic errors to be corrected can drift slowly over time. Thus, the error estimation needs to be performed on time scales smaller than the drift time in order to correct the errors before the estimates become obsolete. However, as adaptive techniques can take advantage of experimental data collected at each measurement step, they can be successfully employed for increasing the information obtained with each measurement and thus decreasing the amount of resources needed as compared to non-adaptive techniques.



Along this line, in this work, we propose and explore an adaptive method based on the application of Bayesian inference for the characterisation of medium-scale multi-qubit states. For concreteness, we focus on the estimation of phases appearing in stabiliser states used in quantum error correction. We examine the efficiency of the method by studying the number of measurements needed and we derive an analytical rule to obtain the optimal measurement to perform at each time. This simple rule does not rely on numerical calculations and ensures the adaptiveness of the method  to find the optimal measurement at each step. This renders the protocol particularly suitable for on-chip processing in adaptive control systems. 

In addition, we evaluate the efficiency of our adaptive method and compare it to that of the Phase Optimisation Method (PHOM), a non-adaptive phase estimation method developed in \cite{muller2016iterative}. The latter method is based on a generalization of a Ramsey experiment to determine and compensate systematic phases appearing in multi-qubit states and it was proposed and performed experimentally for the seven-qubit quantum error correcting color code (Steane code) \cite{steane1996error}. For this reason, we choose to evaluate the efficiency of our adaptive method for the phase characterisation of the states used for the Steane code and we show that our method has an improvement in the efficiency compared to the PHOM.


Remarkably, this adaptive method is not restricted only to the Steane code, but can be used for other multi-qubit states since it only relies on the application of simple single-qubit operations and measurements. As a consequence, the results shown in this paper are applicable to the optimisation of other QEC codes and the compensation of systematic errors appearing in other physical platforms for quantum-information processing such as, e.g, trapped ions \cite{haffner2008quantum,nigg2014quantum,sriarunothai2018speeding,johanning2009quantum}, Rydberg atoms \cite{jaksch2000fast,saffman2010quantum,crow2016numerical} in optical lattices \cite{anderson2011trapping,viteau2011rydberg,schauss2012observation} or tweezer arrays \cite{nogrette2014single,xia2015randomized} or other AMO or solid-state architectures \cite{hanson2008coherent,corcoles2015demonstration,gambetta2017building,kelly2015state,waldherr2014quantum,fedorov2012implementation}.

This paper is organised as follows: In Section II we introduce the concepts and notation for a one-qubit state phase measurement by using a Ramsey experiment and a Bayesian inference process. In Section III we briefly review basic properties of the Steane code to which we will apply our technique. In Section IV we compare the efficiencies of the PHOM proposed in Ref. \cite{muller2016iterative} and our Bayesian inference method to an intermediate quantum state obtained during the preparation of the logical states of the Steane code since this intermediate state has a less complex structure than the final states of the code. In Section V we generalise the previous results and present a Bayesian inference method to estimate the phase shifts on the fully encoded seven-qubit logical states. Finally, in Section VI we summarise our results, especially the comparison between the results obtained for the efficiency of our method with the method in Ref. \cite{muller2016iterative}, and conclude with a brief outlook.

\section{One qubit case}
\begin{figure}[h]
	\begin{center}
		\includegraphics[width=0.95\textwidth]{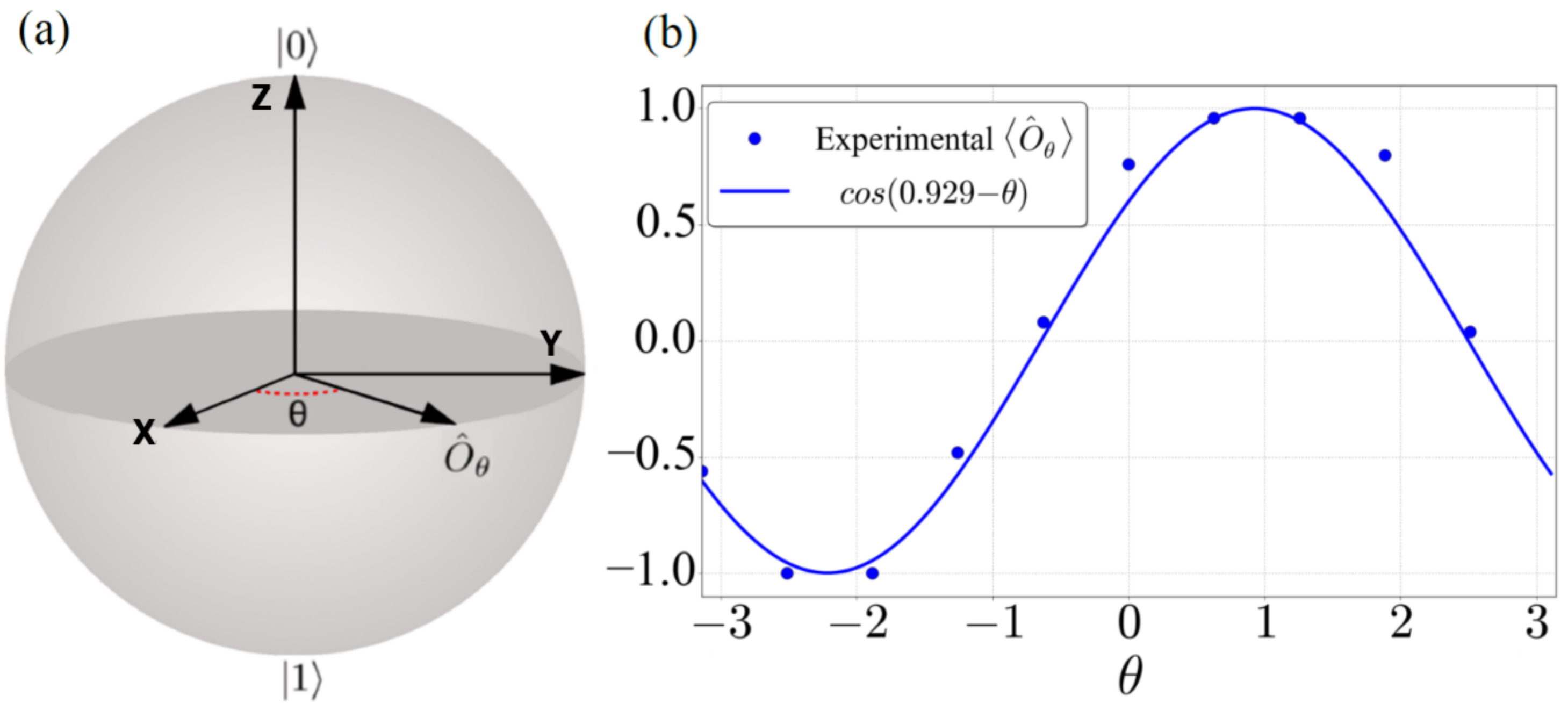}
		\captionsetup{width=0.95\linewidth, labelfont=bf}
		\caption{(a) Bloch sphere representation of the measurements to be performed for estimating the phase $\phi$ of the quantum state of Eq.~\eqref{qubitstate}. The observables $\hat{O}_\theta$ are obtained by rotating the observable $\hat{X}$ around the $Z$ axis by an angle $\theta$. The way $\hat{O}_\theta$ is selected depends on the method used to estimate $\phi$. In the Ramsey scan case different values of $\theta$ are selected and several measurements are performed for each of these values. This allows a reconstruction of the expected sinusoidal dependence of $\bra{\psi}\hat{O}_\theta\ket{\psi}$ with $\theta$ that yields an estimate for the phase $\phi$ [panel (b)]. In the Bayes case, the outcomes of $\hat{O}_\theta$ are used to update the probability distribution $P(\phi)$. In this case, the value $\theta$ for each measurement is different and it is selected in a way that maximises the information gain per measurement (See Sec~\ref{BayesInferenceSec} and Fig.~\ref{bigbig}). (b) Ramsey scan simulation for the phase estimation of the state $\ket\psi$ in Eq.~\eqref{qubitstate} with $\phi=1$. The points represent the values $\hat{O}_{\theta_m}$ where the $\theta_m$ are $M=10$ equidistant points in the interval $[-\pi,\pi)$. The points are obtained by simulating $N=50$ measurements of $\hat{O}_{\theta_m}$ for each $\theta_m$  and they are used to fit a cosine whose phase is the estimation obtained for $\phi$.}
		\label{Blochsphere}
	\end{center}
\end{figure}

In this Section, we will show how to estimate the unknown phase $\phi$ of the following quantum state
\begin{equation}
\label{qubitstate}
\ket{\psi}=\frac{1}{\sqrt{2}}\left( \ket{0}+e^{i\phi}\ket{1}\right)
\end{equation}
from a finite set of data obtained from measurements. We suppose we can prepare as many copies of $\ket{\psi}$ as needed. Since measurements of the $\hat Z$ Pauli operator yield no information about $\phi$ we will perform measurements of the operator
\begin{equation}
\label{thetaoperator}
\hat O_\theta=\cos(\theta)\hat X + \sin(\theta)\hat Y
\end{equation}
on the XY plane of the Bloch sphere (see Fig.~\ref{Blochsphere} (a)). Thus, the expected value of this operator for the state $\ket{\psi}$ is
\begin{equation}
\label{expectedpsi}
\bra{\psi}\hat O_\theta\ket{\psi}= \cos(\phi-\theta).
\end{equation}
In the following, we will study two different ways in which we can select $\theta$ in Eq.~\eqref{thetaoperator} in order to obtain the value $\phi$: Ramsey scan and a Bayesian inference process.

\subsection{Ramsey scan}\label{RamseySec}
In order to estimate $\phi$, one can apply a  Ramsey-type experiment that can be summarised as follows (see Fig.~\ref{Blochsphere}): First we divide the interval $[-\pi,\pi)$ in $M$ equidistant points
\begin{equation}
\theta_m = m\cdot\frac{2\pi}{M}-\pi,\quad m=0,1,...,M-1 \ .
\end{equation}
For each of the values $\theta_m$ we estimate the expected value of $\hat O_{\theta_m}$ by $
\label{meanO}
\bra{\psi} \hat O_{\theta_m} \ket{\psi}= ({N^{(m)}_+ - N^{(m)}_-})/{N}
$
where $N^{(m)}_+$ ($N^{(m)}_-$) is the number of times we obtain a $+$ ($-$) when measuring in the $\hat O_{\theta_m}$ basis and $N$ is the total number of measurements. Using the values $\bra{\psi} \hat O_{\theta_m} \ket{\psi}$ obtained, we perform a least squares fit to a function of the form
\begin{equation}
\label{fit}
f(\theta)=\cos(A-\theta)
\end{equation}
where $A$ is a phase. By comparison of Eqs.~\eqref{expectedpsi} and \eqref{fit}, our estimate of $\phi$ is given by the fitted parameter $A$.

\subsection{Bayesian inference}\label{BayesInferenceSec}

\begin{figure}
		\begin{center}
			\includegraphics[width=0.97\textwidth]{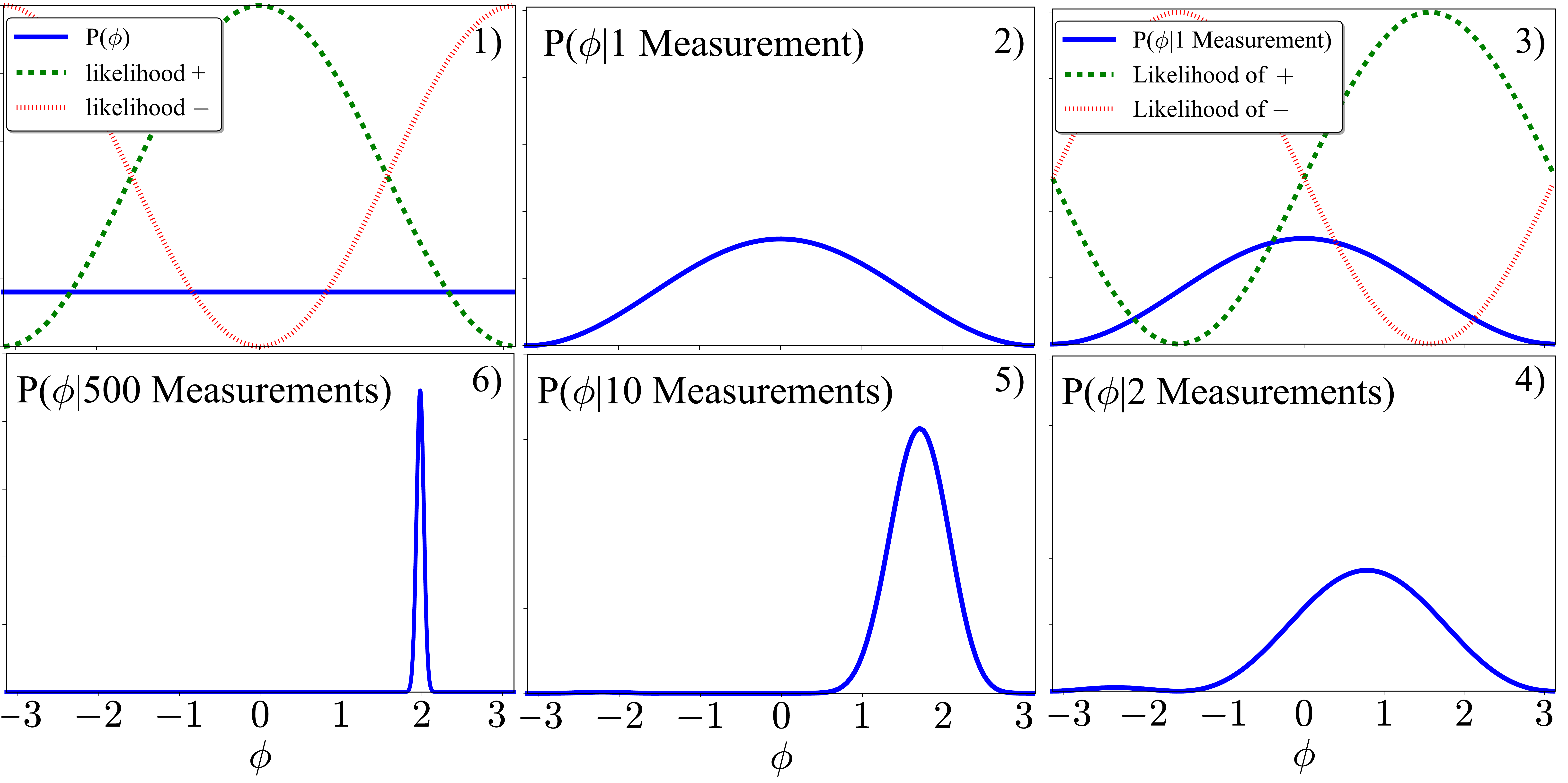}
			\captionsetup{width=0.95\linewidth, labelfont=bf}
			\caption{Bayesian inference process for the phase estimation of the state $\ket{\psi}$ in Eq.~\eqref{qubitstate} with $\phi=2$. Each step of the inference process consists in performing a measurement of the operator $\hat{O}_{\theta_{n-1}}$, updating the probability distribution based on the result of the measurement and selecting an optimal angle $\theta_n$ for the next measurement. 1) One starts with a uniform probability distribution of $\phi$. $\theta=0$ is selected for the first measurement. 2) If the $+$ outcome is obtained (as assumed here), the probability distribution is updated by multiplying the prior probability distribution by the likelihood of obtaining a $+$ and renormalizing. The maximum of $P(\phi|1 \text{\ measurement})$ is located at $\phi=0$. 3) The optimal selection of $\theta$ will be given by Eq.~\eqref{optimaltheta}. Thus, we select $\theta=\pi/2$ for the next measurement. 4) Assume $+$ is obtained again. The probability distribution can be updated again based on this result. This process of measuring, updating and finding the next optimal $\theta$ can be performed iteratively. 5) After $10$ measurements the probability distribution can be approximated by a normal distribution. 6) After $500$ measurements we obtain a normal distribution centered near the value $\phi=2$ used for the simulation.}
			\label{bigbig}
		
		\end{center}
	\end{figure}

In this section we discuss how to use Bayes' theorem in order to estimate the value of $\phi$.  Bayes' theorem  prescribes how to update the prior probability distribution of $\phi$, $P(\phi)$, after a measurement $\mathcal{M}$ using its likelihood, $P(\mathcal{M}|\phi)$ (see Fig.~\ref{bigbig}). The final result is a posterior probability distribution of $\phi$, $P(\phi|\mathcal{M})$, with the form
\begin{equation}
\label{bayes}
P(\phi|\mathcal{M})\propto P(\mathcal{M}|\phi)P(\phi)
\end{equation}
up to a normalization factor. If we perform a new measurement we can apply Bayes' theorem again and use the obtained posterior as a prior for the next measurement and so on. For an increasing number of measurements the degree of uncertainty of $\phi$ will decrease, allowing us to reach a desired value in the uncertainty of the estimated value of $\phi$. In our case we measure the operators $\hat O_\theta$ (with different values of $\theta$ for each measurement) with possible outcomes $+_\theta$ and $-_\theta$. The likelihoods of these outcomes for the state as given by Eq.~\eqref{qubitstate} are
\begin{equation}\label{eqn_one_qubit_likelihood}
P(\pm_\theta | \phi)=\frac{1\pm \cos(\phi-\theta)}{2}.
\end{equation}
Assuming no prior knowledge about the value of $\phi$ we start with a uniform probability distribution $P(\phi)={1}/{2\pi}$ as a prior. After $N$ measurements and applying Eq.~\eqref{bayes} iteratively, the probability distribution for $\phi$ is given by
\begin{align}\label{eqn:posterior}
P(\phi|\pm_{\theta_{1}},...,\pm_{\theta_{N}})\propto P(\phi) P(\pm_{\theta_{1}}|\phi)\cdot ... \cdot P(\pm_{\theta_{N}}|\phi)= \nonumber\\
=\frac{1}{2\pi}\cdot \frac{1\pm \cos(\phi-\theta_{1})}{2}\cdot ... \cdot \frac{1\pm \cos(\phi-\theta_{N})}{2}.
\end{align}
As the number of measurements increases, the posterior probability distribution can be approximated by a normal distribution with decreasing standard deviation. 

\subsubsection{Efficiency of the parameter learning process.}

\begin{figure}
	\begin{center}
		\includegraphics[width=0.95\textwidth]{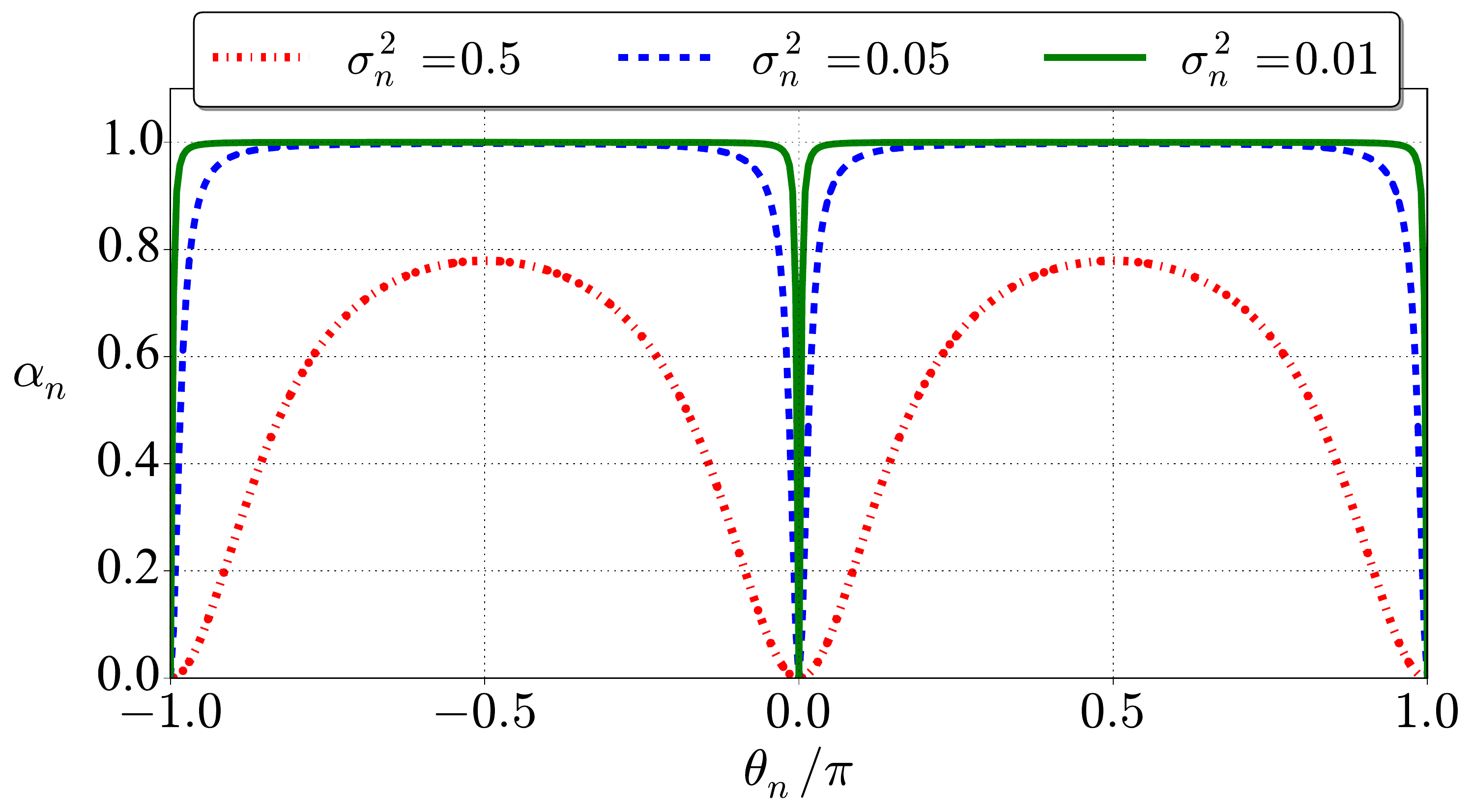}
		\captionsetup{width=0.95\linewidth, labelfont=bf}
		\caption{The quantity $\alpha$ dictates the decrease of the variance $\overline{\sigma^2_{n+1}-\sigma^2_n}$ after each measurement step. This figure shows the dependence of $\alpha$ with $\theta_n$ for the case $\bar{\phi}_n=0$ and different values of $\sigma^2_n$. Other values of $\bar{\phi}_n$ produce the same plot with a translation on the horizontal axis. As $\sigma^2_n$ decreases $\alpha$ approaches $1$ except for the values $\theta=\bar{\phi}_n\pm k\pi$.}
		\label{alphafactor1q}
	\end{center}
\end{figure}

In this section we show how the behaviour of the standard deviation of the distribution in Eq.~\eqref{eqn:posterior} allows us to choose the value $\theta$ after each measurement in order to maximize the information gain of $\phi$. It is expected that as we perform more measurements, the mean value of the probability distribution  Eq.~\eqref{eqn:posterior} gets closer to the true value of $\phi$ and its standard deviation decreases. Let us suppose that, after a sufficiently large number $n$ of measurements, the probability distribution $P_n(\phi)$  can be approximated by a Gaussian with mean $\bar\phi_n$ and standard deviation $\sigma_n$. For the measurement $n$, the probability $p^\pm_{n}$ of measuring $\pm$ when the angle selected is ${\theta_n}$ is
\begin{equation}
p^\pm_{n}=\int_{-\pi}^{\pi}  \frac{1\pm\cos(\phi-\theta_{n})}{2} P_n(\phi) d\phi\ .
\end{equation}
The probability distribution after having obtained $+$ or $-$ is updated as
\begin{equation}
P^\pm_{n+1}(\phi)=\frac{1}{p^\pm_n}\frac{1\pm\cos(\phi-\theta_{n})}{2} P_n(\phi),
\end{equation}
where $p^\pm_n$ appears due to normalization. These posterior probability distributions will have a standard deviation denoted by $\sigma^\pm_{n+1}$. We obtain that the average decrease of the variance after performing measurement $n$ is
\begin{equation}
\label{variancebehaviour}
\overline{\sigma^2_{n+1}-\sigma^2_n}=-\alpha_n\sigma^4_n
\end{equation}
with
\begin{equation}
\label{alphadefinition}
\alpha_n \equiv \frac{e^{\sigma^2_n}\sin^2(\overline\phi_n-\theta_n)}{1-e^{\sigma^2_n}\cos^2(\overline\phi_n-\theta_n)}\ .
\end{equation}
From inspection of Eqs.~\eqref{variancebehaviour} and \eqref{alphadefinition}, we conclude that the maximum decrease on average for the variance is obtained when we select a value of $\theta_n$ that maximizes the value of $\alpha_n$. This is achieved (see Fig.~\ref{alphafactor1q}) for
\begin{equation}
\label{optimaltheta}
\theta_n=\overline{\phi}_n\pm\frac{\pi}{2}.
\end{equation}
For this selection, the value of $\alpha_n$ approaches a constant value. Then, it can be proven that the succession in Eq.~\eqref{variancebehaviour} has the following asymptotic solution:
\begin{equation}
\label{alphasd}
\sigma^2_n=\frac{1}{\alpha_n n}.
\end{equation}
However, as the variance decreases, $\alpha_n$ approaches the constant value~$1$ (except for values close to $\theta_n=\overline{\phi}_n\pm k\pi$, $k\in \mathbb{N}$) as Fig.~\ref{alphafactor1q} shows. This means that after several measurements, the decrease of the variance will be independent on  the value of $\theta$ we select for the next measurement and $\sigma^2_n$ will evolve as
\begin{equation}
\label{sql}
\sigma^2_n=\frac{1}{n}.
\end{equation}
Similar results can be obtained by performing calculations involving Shannon's entropy \cite{negnevitsky2018feedback} for the selection of the optimal measurement. However, this approach requires numerical calculations to obtain the optimal measurement in each step of the iteration. This increases the computational resources needed for in-situ optimisation between experimental measurement runs as compared to the simple rule governed by Eq.~\eqref{optimaltheta} of our method.
It is worth mentioning that the result in Eq.~\eqref{sql} indicates that our increase in the knowledge of the system satisfies the Standard Quantum Limit (SQL). Although there are methods that aim to obtain better results than the ones given by this limit \cite{wiebe2016efficient,higgins2007entanglement,giovannetti2004quantum,paesani2017experimental}, these methods rely on having access to, e.g., entangled states, which we will not consider in this work.

\section{Characterisation of multi-qubit states}

In quantum error correction the quantum information is encoded in entangled many-qubit systems. This provides protection against noise.

\begin{figure}[ht]
	\begin{center}
		
		\includegraphics[width=0.7\textwidth]{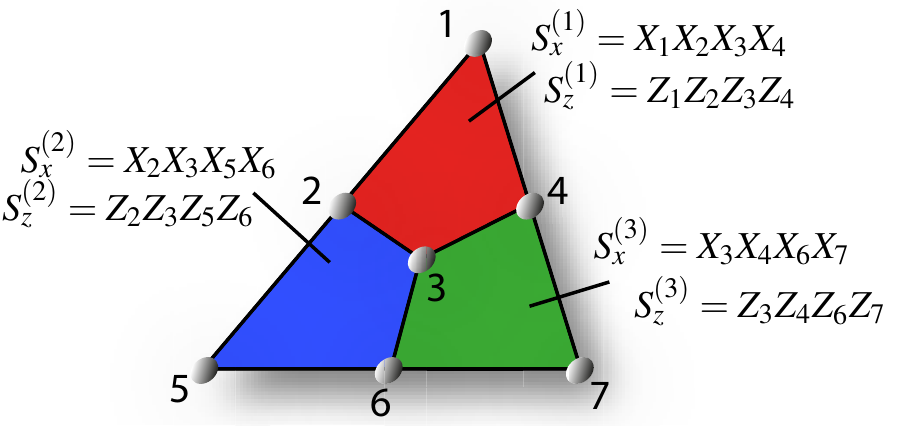}
		\captionsetup{width=0.95\linewidth, labelfont=bf}
		\caption{Steane code or seven-qubit color code:  A structure composed of seven qubits is used for encoding one logical qubit. The code is defined by six stabiliser generators $S^{(j)}_X$ and $S^{(j)}_Z$ for $j=1,2,3$ associated with each of the four-qubit plaquettes. The code space is defined as the simultaneous $+1$ eigenspace of these stabilisers.}
		\label{steanecode}
	\end{center}
\end{figure}

In the following sections we will discuss the estimation of systematic errors appearing in the preparation of the stabiliser states used in quantum error correction. For concreteness, we focus on the states used for the seven-qubit Steane code \cite{steane1996error}. This code represents the minimal instance of 2D color codes \cite{bombin2006topological} and it is obtained by restricting a the Hilbert space of seven qubits to the subspace of states which are simultaneous $+1$ eigenstates of six commuting stabiliser operators $S^{(i)}_x$ and $S^{(i)}_z$, $i=1,2,3$ (see Fig.~\ref{steanecode}). These stabilisers define a 2-dimensional subspace for this 7-qubit system that can be used to encode a logical qubit. Additionally, the logical $X$ and $Z$ operators can be chosen as $X_L=\prod_{i=1}^{7}X_i$ and $Z_L=\prod_{i=1}^{7}Z_i$. The logical state $\ket{0}_L$ is defined by $Z_L\ket{0}_L=\ket{0}_L$,
\begin{align}
\ket{0}_L\propto (1+Z_L)\prod_{i=1}^{3} (1+S^{(i)}_x)\ket{0}^{\otimes 7}.
\end{align}
Similarly, the  logical $\ket{1}_L$ is defined by $Z_L\ket{1}_L=-\ket{1}_L$. These states satisfy $\ket{1}_L=X_L\ket{0}_L$ and $\ket{0}_L=X_L\ket{1}_L$.

In the following sections we develop a method to measure the phases appearing due to systematic errors in the preparation of this class of states. For simplicity, we will first study an intermediate case of the full 7-qubit Steane encoding process (two-plaquette case) to introduce the concepts that will be needed to correct the phases appearing in the fully encoded system (three-plaquette case).

\section{Two-plaquette case}
\label{2plaquettesec}

\begin{figure}[ht]
	\begin{center}
		
		\includegraphics[width=0.7\textwidth]{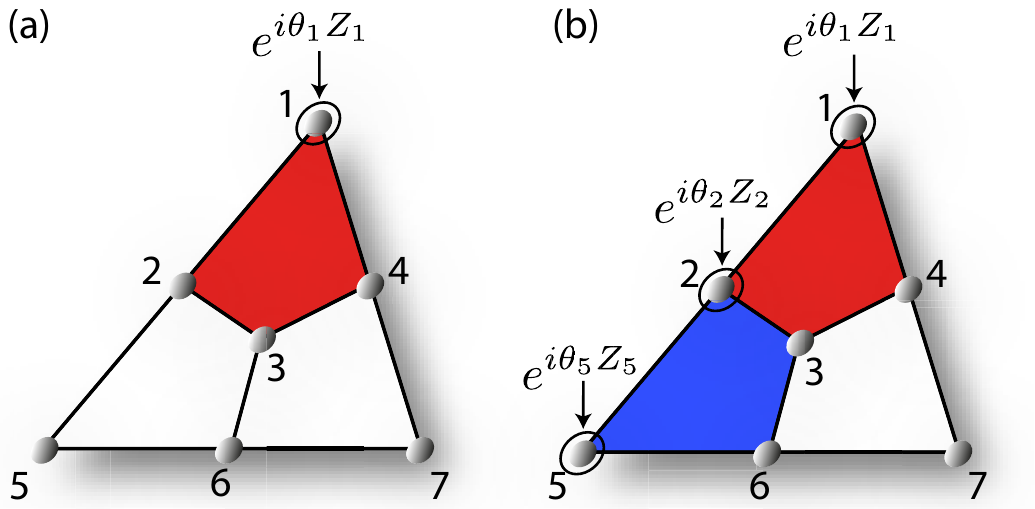}
		\captionsetup{width=0.95\linewidth, labelfont=bf}
		\caption{(a) One-plaquette case. During the first step of the preparation of the seven-qubit error correcting code, four-qubit entangling operations are performed on qubits 1 to 4. A phase appearing in the resulting state due to systematic errors can be estimated by performing measurements of $S^{(1)}_x$ for different rotations on the first qubit. This is similar to the single-qubit phase estimation. (b) Two-plaquette case. After the manipulation of the first and second plaquettes, up to three relative phases can appear in quantum state obtained. To estimate these phases we can perform measurements of $S^{(1)}_x$, $S^{(2)}_x$ and $S^{(1)}_xS^{(2)}_x$ for different rotations on the first, second and fifth qubit.}
		\label{colorcodeplaquettes}
	\end{center}
\end{figure}

At the start of the preparation of the seven-qubit quantum error correcting code, four-qubit entanglement operations are applied to the first plaquette (Fig.~\ref{colorcodeplaquettes} (a)). This yields the quantum state $\ket{\psi_1}\propto (1+S^{(1)}_x)\ket{0}^{\otimes 7}$composed by the superposition of two components in the computational basis which can have a relative phase due to systematic errors.  This is equivalent to a single-qubit phase estimation, as the phase can be corrected by rotating one of the four qubits and performing measurements of the $S^{(1)}_x$ stabiliser.

Therefore, we consider the state $\ket{\psi_2}$ obtained by application of four-qubit entangling operations to the first and second plaquettes (see Fig.~\ref{colorcodeplaquettes} (b))
\begin{equation}\label{eqn_psi_two_plaquette}
\ket{\psi_2}\propto (1+S^{(1)}_x) (1+S^{(2)}_x) \ket{0}^{\otimes 7}. 
\end{equation}
This state maximizes the mean value of the X-type stabilisers on the first and second plaquette, $S^{(1)}_x=X_1X_2X_3X_4$, $S^{(2)}_x=X_2X_3X_5X_6$ and the product of both, $S^{(1)}_xS^{(2)}_x=X_1X_4X_5X_6$:
\begin{equation}
\bra{\psi_2}S^{(1)}_x\ket{\psi_2}=\bra{\psi_2}S^{(2)}_x\ket{\psi_2}=\bra{\psi_2}S^{(1)}_xS^{(2)}_x\ket{\psi_2}=1.
\end{equation}
However, systematic phase shifts accumulate during the preparation of $\ket{\psi_2}$ due to experimental errors. The state $\ket{\psi'_2}$ containing these unknown phase shifts is
\begin{equation}
\label{systematicerrorstate}
 \ket{\psi'_2}  \propto \ket{0000000}+e^{i\phi_1}\ket{0110110}+e^{i\phi_2}\ket{1111000}+e^{i\phi_3}\ket{1001110}.
\end{equation}
In order to compensate these relative phase shifts we can apply single qubit Z rotations (see Fig.~\ref{colorcodeplaquettes} (b)). For example, by rotating the first, second and fifth qubits we obtain
\begin{align}
e^{i\theta_1Z_1}e^{i\theta_2Z_2}e^{i\theta_5Z_5}\ket{\psi'_2} \propto \ket{0000000}+e^{i[\phi_1+2(\theta_2+\theta_5)]}\ket{0110110} \nonumber\\+e^{i[\phi_2+2(\theta_1+\theta_2)]}\ket{1111000}+e^{i[\phi_3+2(\theta_1+\theta_5)]}\ket{1001110}.
\end{align}

The selection of qubits to be rotated is arbitrary as long as these three rotations do not commute with $S^{(1)}_x$, $S^{(2)}_x$ and $S^{(1)}_xS^{(2)}_x$, respectively. The expected values of the stabilisers in the state Eq.~\eqref{systematicerrorstate} are given by
\begin{align}
\langle S^{(1)}_x\rangle =\frac{\cos[\phi_2+2(\theta_2+\theta_1)]+\cos[\phi_1-\phi_3+2(\theta_2-\theta_1)]}{2}, \label{eqn_stabilizerS1} \\
\langle S^{(2)}_x\rangle =\frac{\cos[\phi_1+2(\theta_2+\theta_5)]+\cos[\phi_2-\phi_3+2(\theta_2-\theta_5)]}{2}, \\ \langle S^{(1)}_xS^{(2)}_x\rangle =\frac{\cos[\phi_3+2(\theta_5+\theta_1)]+\cos[\phi_1-\phi_2+2(\theta_5-\theta_1)]}{2},
\end{align}
In order to obtain information about the unknown systematic phases, we can perform measurements of these stabilisers for different values of the rotation angles $\theta$. Once the values of these phases are measured it is possible to perform single-qubit rotations to transform the state $\ket{\psi'_2}$ into the desired state $\ket{\psi_2}$. A way to obtain these values is the Phase Optimisation Method \cite{muller2016iterative}. We propose another method based on application of Bayesian inference. In the following subsections we review the PHOM and introduce our Bayesian protocol.

\subsection{Phase Optimisation Method}\label{PHOM}

The Phase Optimisation Method introduced in Ref. \cite{muller2016iterative} is given by the following iterative protocol. For concreteness, here we review how it works for the optimisation of the state in Eq.~\eqref{systematicerrorstate}.
\begin{enumerate}
	\item \label{PHOM1} For each stabiliser, an associated rotation on a qubit $i$, $\theta_i$, is chosen. The selection is arbitrary, but each stabiliser must not commute with its associated qubit rotation. We associate $\theta_2$ with $S^{(1)}_x$, $\theta_5$ with $S^{(2)}_x$ and $\theta_1$ with $S^{(1)}_xS^{(2)}_x$.
	\item \label{PHOM2} Choose an initial configuration for the set of rotation parameters $\boldsymbol{\theta}^{(0)}=\set{\theta_1^{(0)}, \theta_2^{(0)}, \theta_5^{(0)}}$.
	\item \label{PHOM3} Scan $\langle S^{(1)}_x\rangle$ in a similar way as in the single-qubit case (see Fig.~\ref{Blochsphere} (b)) over its associated angle, $\theta_2$, in the interval $[-\pi,\pi]$ while keeping $\theta_1=\theta^{(0)}_1$ and $\theta_5=\theta^{(0)}_5$ fixed. Determine and fix $\theta_2$ to the value $\theta_2=\theta^{(1)}_2$ for which $\langle S^{(1)}_x\rangle$ is maximized. Similarly perform scans of $\langle S^{(2)}_x\rangle$ over $\theta_5$ to obtain $\theta^{(1)}_5$ and $\langle S^{(1)}_xS^{(2)}_x\rangle$ over $\theta_1$ to obtain $\theta^{(1)}_1$. With these steps $\boldsymbol{\theta}^{(0)}=\set{\theta_1^{(0)}, \theta_2^{(0)}, \theta_5^{(0)}}$ has changed to $\boldsymbol{\theta}^{(1)}=\set{\theta_1^{(1)}, \theta_2^{(1)}, \theta_5^{(1)}}$.
	\item \label{PHOM6} The values of the angles $\boldsymbol\theta$ might not converge to the values that correct the phases after only one iteration. Thus, repeat step \eqref{PHOM3} until the set of angles $\boldsymbol\theta$ converges to a desired precision.
\end{enumerate}
This method gives an estimate of the angles $\theta_1$, $\theta_2$ and $\theta_5$ that correct the systematic phases $\phi_1$, $\phi_2$ and $\phi_3$. The precision of this estimate will become better as the number of measurements used for the scans increases. In this work we also introduce a variation of this PHOM to measure phases that we call the constant cosine PHOM.

\subsubsection{Constant cosine PHOM.}

The constant cosine PHOM is a similar method that also performs scans of the expected values of the stabilisers for different qubit rotations to obtain a correction for systematic phases. This process is more similar to a Ramsey experiment as only one scan of each stabiliser is needed. From  Eq.~\eqref{eqn_stabilizerS1} we can see that, if  we  keep the value $\theta_2-\theta_1$ fixed and vary $\theta_2+\theta_1$, the mean value of  $\langle S^{(1)}_x\rangle$  will be given by
\begin{equation}\label{eqn_constant_cos_phom}
\langle S^{(1)}_x\rangle =\frac{\cos[\phi_2+2(\theta_2+\theta_1)]+h}{2}, \quad h\equiv \cos[\phi_1-\phi_3+2(\theta_2-\theta_1)],
\end{equation}
where $h$ is constant for all the measurements since the difference $\theta_2-\theta_1$ is fixed. Thus the angle $\phi_2$ that represents the phase shift to be corrected is given by the value of $-2(\theta_2+\theta_1)$ for which a maximum in the mean value $\langle S^{(1)}_x\rangle$ is reached. By analysing $\langle S^{(2)}_x\rangle$ and $\langle S^{(1)}_xS^{(2)}_x\rangle$ in a similar way one can then obtain $\phi_1$ and $\phi_3$, too.

\subsection{Bayesian inference method}

In the following we will introduce and analyse two Bayesian inference methods to measure the phases in the state $\lvert\psi'_2\rangle$ of Eq.~\eqref{systematicerrorstate} of the two-plaquette case. These methods are (i) a Bayesian inference method by direct application of the likelihoods, and (ii) a Bayesian inference method using marginal likelihoods. We will first describe this direct Bayesian inference method and then explain the method we propose to improve the PHOM, namely the marginal likelihood method.

\subsubsection{Direct Bayesian inference method.}
An estimation of the phases using Bayesian inference is performed by measuring the plaquettes and updating the probability distribution based on the results obtained. The likelihoods for each plaquette measurement can be obtained from the expressions for the expected values by
\begin{equation}
P_1(\pm_{\boldsymbol{\theta}}|\boldsymbol{\phi})=\frac{1\pm\langle S^{(1)}_x\rangle}{2},\quad P_2(\pm_{\boldsymbol{\theta}}|\boldsymbol{\phi})=\frac{1\pm\langle S^{(2)}_x\rangle}{2},\quad P_{12}(\pm_{\boldsymbol{\theta}}|\boldsymbol{\phi})=\frac{1\pm\langle S^{(1)}_x S^{(2)}_x\rangle}{2},
\end{equation}
where $P_1(\pm_{\boldsymbol{\theta}} |\boldsymbol{\phi})$ is the likelihood of obtaining a $+$ or a $-$ outcome when measuring $S^{(1)}_x$ for $\boldsymbol{\theta}=\{\theta_1, \theta_2, \theta_5\}$ and $\boldsymbol{\phi}=\{\phi_1, \phi_2, \phi_3\}$. Similarly, $P_2(\pm_{\boldsymbol{\theta}} |\boldsymbol{\phi})$ is related to $S^{(2)}_x$ and $P_{12}(\pm_{\boldsymbol{\theta}} |\boldsymbol{\phi})$ to $S^{(1)}_xS^{(2)}_x$. For instance, the likelihood for $S^{(1)}_x$ is given by 
\begin{align}
P_1(\pm_{\boldsymbol{\theta}} |\boldsymbol{\phi})=\frac{2\pm\cos[\phi_2+2(\theta_2+\theta_1)]\pm \cos[\phi_1-\phi_3+2(\theta_2-\theta_1)]}{4}. 
\end{align}
Since the likelihoods used are functions of three variables, the obtained probability distribution will be a three-dimensional function. In general, if the number of unknown parameters appearing in the likelihoods of the experiment increases, the probability distribution obtained will be a function of many variables and, therefore, finding the most probable values for the phases and their variance becomes more difficult. This complication can be avoided if in the measurement of each stabiliser we keep one of the cosines constant in a similar way as it is done for the constant cosine PHOM. This yields a likelihood given by (see also Eq.\eqref{eqn_constant_cos_phom})
\begin{equation}
\label{constcosinelikelihood}
P_1(\pm_{\boldsymbol{\theta}} |\phi_2,h) =\frac{2\pm h_2\pm\cos[\phi_2+2(\theta_2+\theta_1)]}{4}
\end{equation}
for the first plaquette, where the value $\theta_2-\theta_1$ is kept constant to ensure that one of the cosines has a constant value given by $h_2$. Similar expressions can be obtained for the other stabilisers. This approach yields normal probability distributions  defined on two variables, one being $h_1$, $h_2$ or $h_3$ and the other being $\phi_1$, $\phi_2$ or $\phi_3$ depending on the stabiliser that is measured. The estimate for each phase is easily obtained from its corresponding probability distribution.

\begin{figure}[ht]
	\begin{center}
		
		\includegraphics[width=0.9\textwidth]{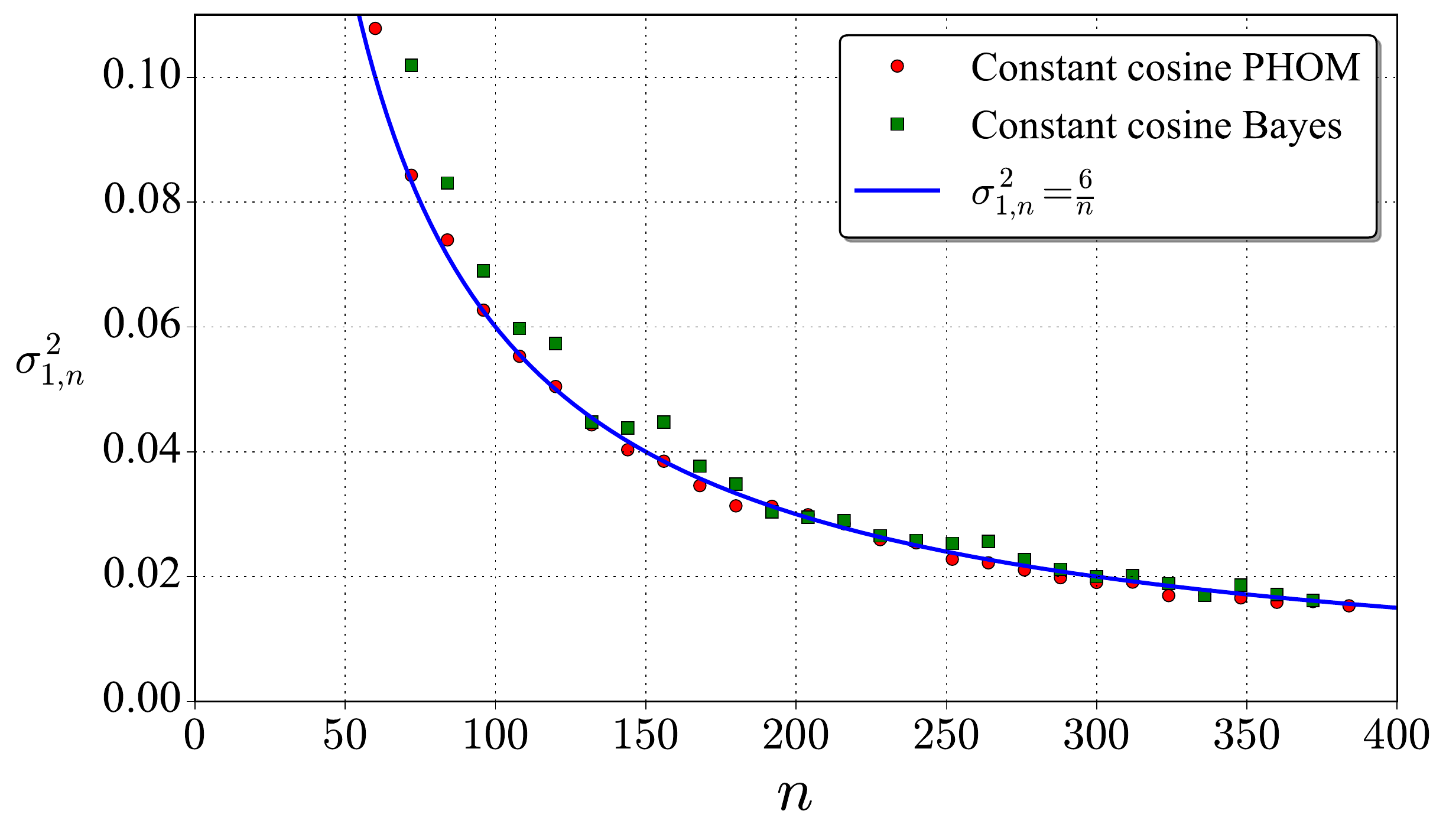}
		\captionsetup{width=0.95\linewidth, labelfont=bf}
		\caption{Behaviour of the variance of $\phi_1$ as a function of the number of measurements $n$ obtained for the simulations of the constant cosine approach for both the PHOM (red circles) and the Bayesian inference (green squares). The solid line represents a numerical fit given by $\sigma^2_{1,n}=6/n$. The same behaviour is obtained for $\phi_2$ and $\phi_3$ (not shown).}
		\label{ConstantCosineSDComparison}
	\end{center}
\end{figure}

We now compare this method and the constant cosine PHOM by simulating them and, by fitting the numerical data we obtained, we find that for both, the scaling of the variance as a function of the number of measurements $n$ is given by  $\sigma^{2}_{i,n}=6/n$ when estimating the single angle $\phi_i$ (see Fig.~\ref{ConstantCosineSDComparison}). However, in order to obtain an estimate of the other two phases, this process must be repeated for each of the other stabilisers. This yields a scaling for the  variance $\sigma^2_{n}$ that scales as
\begin{equation}
\label{sigma2plaq}
\sigma^2_n=\frac{18}{n}.
\end{equation}
that gives an estimate of the efficiency of the PHOM and the direct Bayesian inference method for the intermediate state $\lvert\psi'_2\rangle$.

In the following, we will introduce and analyze  the  Bayesian marginal likelihood method, which constitutes an  improvement in the efficiency both of the PHOM  and of the direct Bayesian inference technique.
\subsubsection{Marginal likelihood method.}

\begin{figure}
	\begin{center}		
		\includegraphics[width=0.9\textwidth]{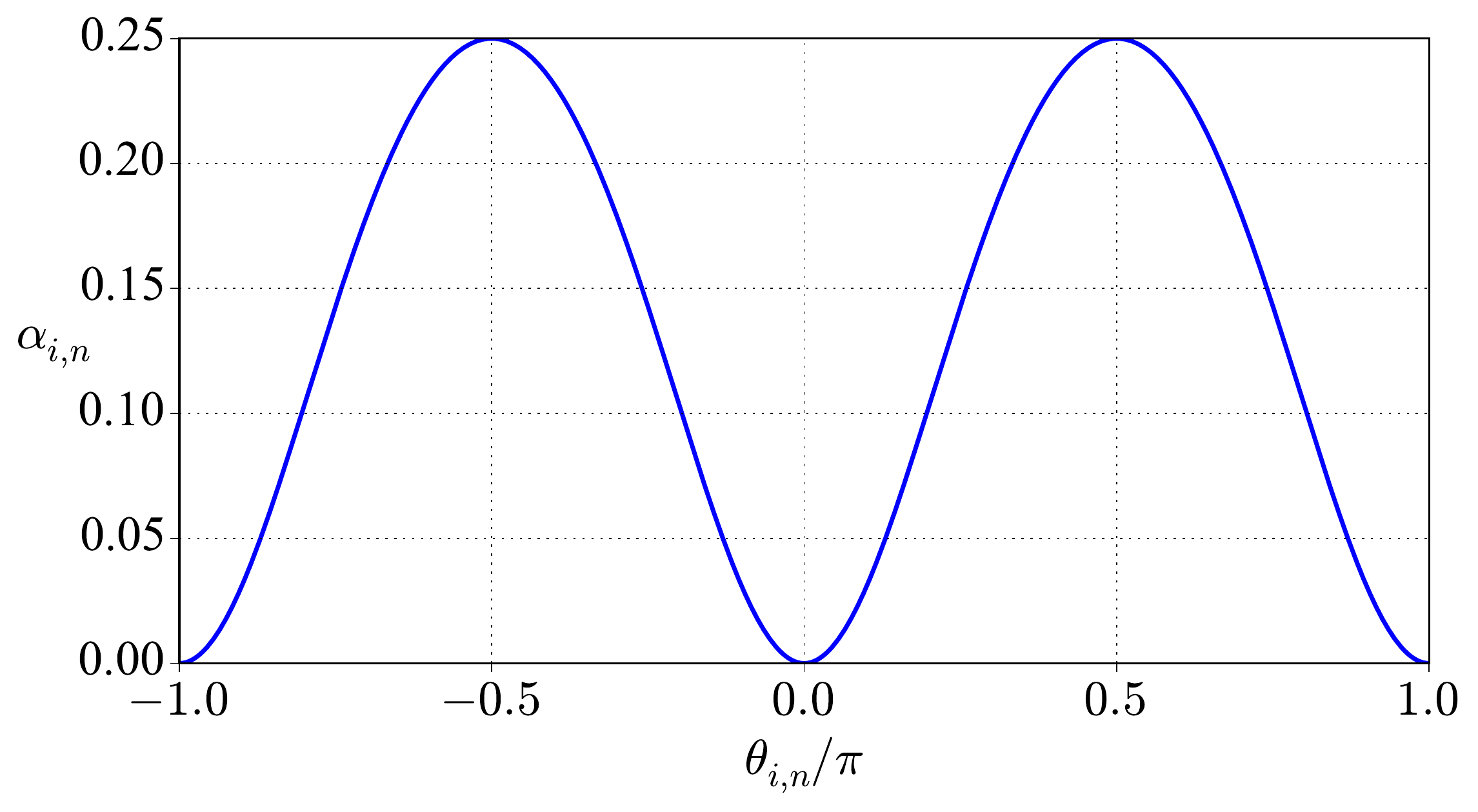}
		\captionsetup{width=0.95\linewidth, labelfont=bf}
		\caption{Dependence of the factor $\alpha_{i,n}$ on $\theta_{i,n}$ for $i=1,2,3$ (here $\bar{\phi}_{i,n}=0$). $\alpha_{i,n}$ dictates the decrease of the variance step by step. $\alpha_{i,n}$ oscillates between $0$ and $0.25$ and its mean value is $(2-\sqrt{3})/2\approx0.134$, with its maximum at $\tilde\theta_{i,n}=\bar{\phi}_{i,n}\pm \pi/2$. By Eq.~\eqref{alphasd}, this is the optimal way of selecting $\theta_n$.}
		\label{alphafactor}
	\end{center}
\end{figure}

\begin{figure}
	\centering
\fbox{\includegraphics[scale=0.325]{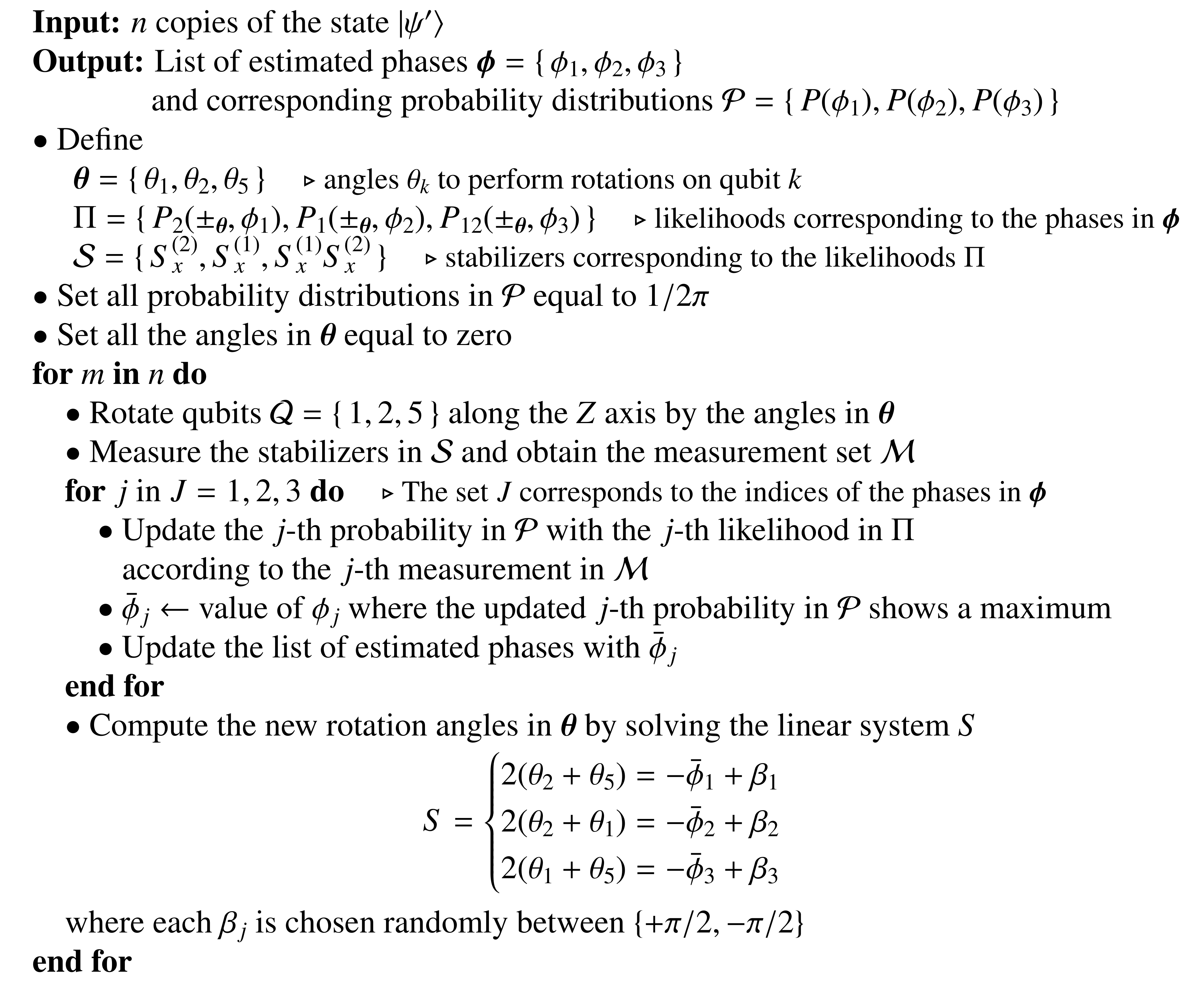}}
		\captionsetup{width=0.95\linewidth, labelfont=bf}
		\caption{Pseudocode to determine the three phases, $\boldsymbol{\phi}=\set{\phi_1,\phi_2,\phi_3}$, appearing in the two-plaquette case of Sec.~\ref{2plaquettesec} by implementation of the marginal likelihood Bayesian inference method. This pseudocode can also be applied for estimating the seven phases appearing in the three-plaquette case of Sec.~\ref{3plaquettesec} after redefining $\boldsymbol{\phi}$, the list of unknown phases and their indices $J$; $\mathcal{P},$ their probability distributions; $\boldsymbol{\theta}$, the angles for the rotations of the measurements; $\Pi$, the likelihoods; $\mathcal{S}$ the list of stabilisers; $\mathcal{Q}$ the list of qubits to be rotated; the system of equations $S$ according to \eqref{generalseven} in \ref{StabilizerAppendix}.}
		\label{pseudocode}
\end{figure}

Let us consider measurements of the first stabiliser whose likelihood is
\begin{equation}
P_1(\pm_{\boldsymbol{\theta}} |\boldsymbol{\phi})=\frac{2\pm\cos[\phi_2+2(\theta_2+\theta_1)]\pm \cos[\phi_1-\phi_3+2(\theta_2-\theta_1)]}{4} \ .
\end{equation}
Suppose we are only interested in $\phi_2$ and we perform measurements with values of $\theta_1$ and $\theta_2$ selected randomly. With this selection the cosine containing $\phi_1-\phi_3$ has a completely random argument and it averages to zero. This yields the following marginal likelihood
\begin{equation}
\label{likelihood2}
P_1(\pm_{\boldsymbol{\theta}} |\phi_2)=\frac{2\pm \cos[\phi_2+2(\theta_2+\theta_1)]}{4},
\end{equation}
which only depends on $\phi_2$. This likelihood is similar to the likelihood for the one qubit case of Eq.~\eqref{eqn_one_qubit_likelihood} so we can generalize the analysis for the scaling of the variance for $\phi_2$ (see Eq.~\eqref{variancebehaviour} and \ref{VarianceBehaviour}) obtaining
\begin{equation}
\overline{\sigma^2_{2,n+1}-\sigma^2_{2,n}} =-\alpha_{2,n}\sigma^4_{2,n},\\
\end{equation}
where
\begin{equation}
\alpha_{2,n} \equiv \frac{\exp({\sigma^2_{2,n}})\sin^2(\overline\phi_{2,n}-\tilde{\theta}_{2,n})}{4-\exp({\sigma^2_{2,n}})\cos^2(\overline\phi_{2,n}-\tilde{\theta}_{2,n})},
\end{equation}
and  $\tilde{\theta}_{2,n}=-2(\theta_{2,n}+\theta_{1,n})$. For small values of $\sigma^2_{2,n}$, $\alpha_{2,n}$ can be approximated by (see Fig.~\ref{alphafactor})
\begin{equation}
\label{alpha2n}
\alpha_{2,n}\approx\frac{\sin^2(\overline\phi_{2,n}-\tilde{\theta}_{2,n})}{4-\cos^2(\overline\phi_{2,n}-\tilde{\theta}_{2,n})}.
\end{equation}
The quantity $\alpha_{2,n}$ oscillates between 0 and $1/4$  and has a mean value equal to $\overline{\alpha}_{2,n}=(2-\sqrt{3})/2\approx0.134$ obtained by integrating Eq.~\eqref{alpha2n} over a uniform probability distribution of $\tilde{\theta}_{2,n}$, since we are using random values of $\tilde{\theta}_{2,n}$. Taking into consideration Eq.~\eqref{alphasd} the scaling of the variance is given by
\begin{equation}
\sigma^2_{2,n}\approx\frac{1}{0.134n}\approx\frac{7.5}{n}.
\end{equation}
A similar derivation can be performed for the likelihoods of the other stabilisers, yielding the same behaviour.
These results are obtained for non-adaptive measurements since so far the experimental configuration used is not selected in a way that maximizes the information gained by each measurement (the $\theta$s are selected randomly). If one chooses $\tilde{\theta}_{i,n}=\bar{\phi}_{i,n}+\pi/2$ for $i=1,2,3$ this causes a displacement $h_i$ to appear in each likelihood, e.g., in the likelihood of $\phi_2$:
\begin{equation}
P_1(\pm_{\boldsymbol{\theta}} |\phi_2,h_2) =\frac{2\pm h_2\pm \cos[\phi_2+2(\theta_2+\theta_1)]}{4}\ . 
\end{equation}
In order to correct this displacement in the likelihoods a random selection between $\tilde{\theta}_{i,n}=\bar{\phi}_{i,n}+\pi/2$ and $\tilde{\theta}_{i,n}=\bar{\phi}_{i,n}-\pi/2$ can be done. As a consequence the displacements will alternate between the value $h_i$ and $-h_i$ and its average will be $0$. Thus, the same likelihood as for the random angles selection case is obtained, but the factor $\alpha_{i,n}$ improves to $1/4$. Thus, the variances scale as $\sigma^2_{i,n}={4}/{n}$. A pseudocode that summarises the steps presented here is shown in Fig.~\ref{pseudocode}.

Using these simple rules for the selection of $\tilde{\theta}_{i,n}$ ensures that the adaptive way of selecting the parameters of the measurements yields more information than selecting them in a non-adaptive way. Additionally, in each measurement one can use the value of all three stabilisers as opposed to what happens in the PHOM and the Bayesian inference with the constant cosine, in which each measurement only yields information about the scanned stabiliser. Thus the estimate of each phase will still have a variance $\sigma^2_n$ given by
\begin{equation}
\sigma^2_n=\frac{4}{n}.
\end{equation}

\section{Three-plaquette case}
\label{3plaquettesec}

The methods seen in the previous sections can be generalized to the more complex case of the entire seven-qubit code. In this case the objective is to measure the 7 phases that can appear in the preparation of the state that represents the logical state $\ket{0}_L$ (see \ref{StabilizerAppendix}). To this end, we can measure seven different combinations of stabilisers: $S^{(1)}_x$, $S^{(2)}_x$, $S^{(3)}_x$, $S^{(1)}_xS^{(2)}_x$, $S^{(1)}_xS^{(3)}_x$, $S^{(2)}_xS^{(3)}_x$ and $S^{(1)}_xS^{(2)}_xS^{(3)}_x$. The marginal likelihood for the measurement of the first stabiliser is
\begin{equation}
P_1(\pm_{\boldsymbol{\theta}} |\phi_2)=\frac{4\pm \cos(\phi_2-\tilde{\theta}_2)}{8},
\end{equation}
where $\tilde{\theta}_2\equiv-2(\theta_1+\theta_2+\theta_3+\theta_4)$. Similar expressions are obtained for the other six combinations of stabilisers (see \ref{StabilizerAppendix}). Following the same process as in the previous sections, the behaviour of the variance and the values of $\alpha$ for each angle $\phi_i$ at the step $n$ ($\alpha_{i,n}$) are now given by (see \ref{VarianceBehaviour})
\begin{equation}
\overline{\sigma^2_{i,n+1}-\sigma^2_{i,n}} = -\alpha_{i,n}\sigma^4_{i,n},
\end{equation}
with
\begin{equation}
\alpha_{i,n}=\frac{e^{\sigma^2_{i,n}}\sin^2(\overline\phi_{i,n}-\tilde{\theta}_{i,n})}{16-e^{\sigma^2_{i,n}}\cos^2(\overline\phi_{i,n}-\tilde{\theta}_{i,n})},
\end{equation}
where for small values of $\sigma^2_{i,n}$ one finds
\begin{equation}
\alpha_{i,n}\approx\frac{\sin^2(\overline\phi_{i,n}-\tilde{\theta}_{i,n})}{16-\cos^2(\overline\phi_{i,n}-\tilde{\theta}_{i,n})}.
\end{equation}
The maximum value of each $\alpha_{i,n}$ is $1/16$ for the selection $\tilde{\theta}_{i,n}=\bar\phi_{i,n}\pm \pi/2$. Thus, this rule ensures an adaptive way of selecting the measurements that yields more information than a non-adaptive selection and gives a scaling as
\begin{equation}
\sigma^2_n=\frac{16}{n}
\end{equation}
for the variance of the estimate for each phase measured.

\begin{figure}[h]
	\begin{center}
		
		\includegraphics[width=\textwidth]{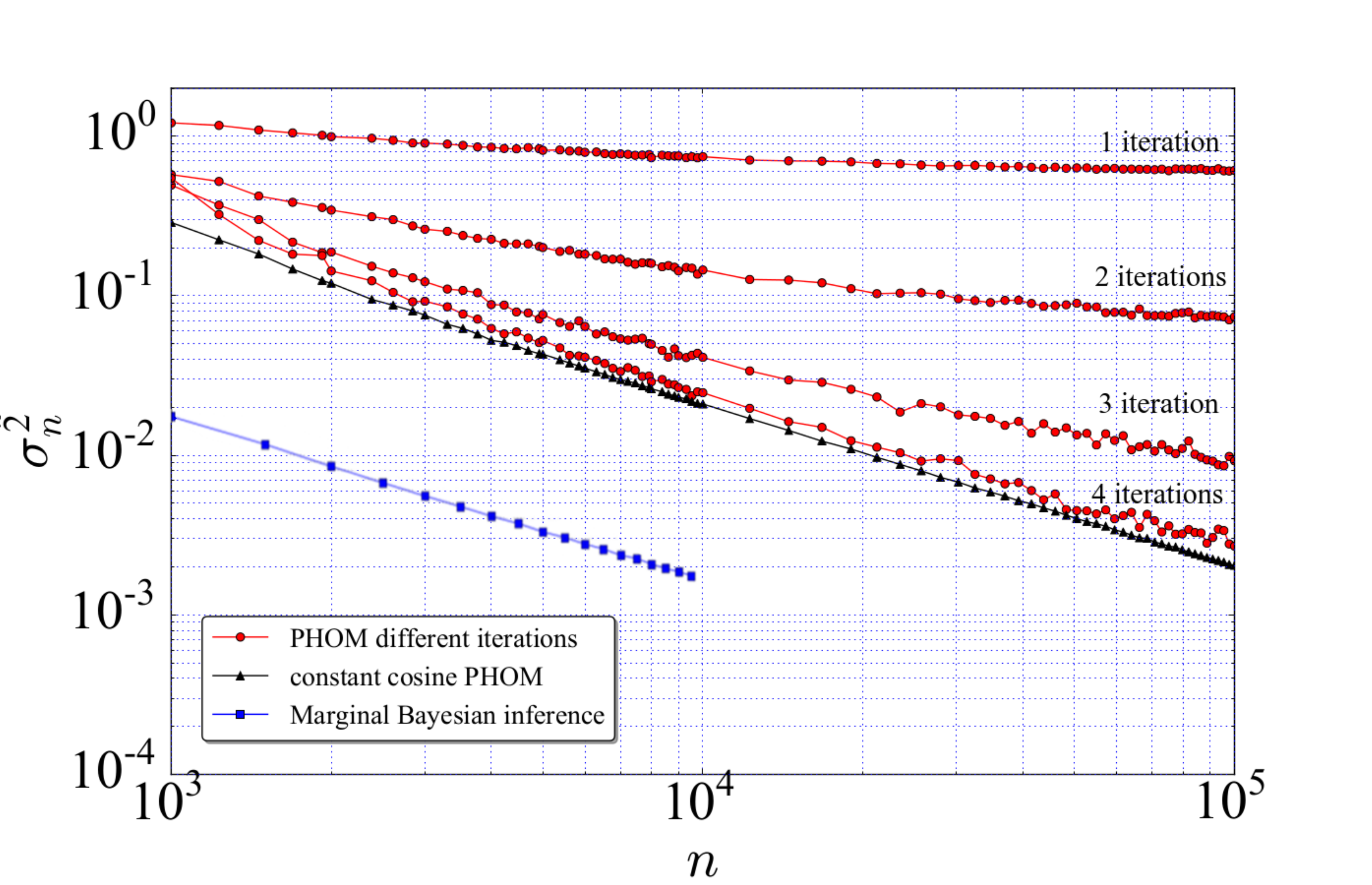}
		\captionsetup{width=0.95\linewidth, labelfont=bf}
		\caption{Behaviour of the variance as a function of the number measurements, $n$, obtained from simulations. For the PHOM using constant cosines a behaviour of $224/n$ is obtained. PHOM shows a similar behaviour when using enough iterations for the method to converge. Finally, using marginal Bayes the performance is $16/n$.}
		\label{Variance_Comparison_3Plaquettes}
	\end{center}
\end{figure}

\begin{figure}[h]
	\begin{center}
		
		\includegraphics[width=0.9\textwidth]{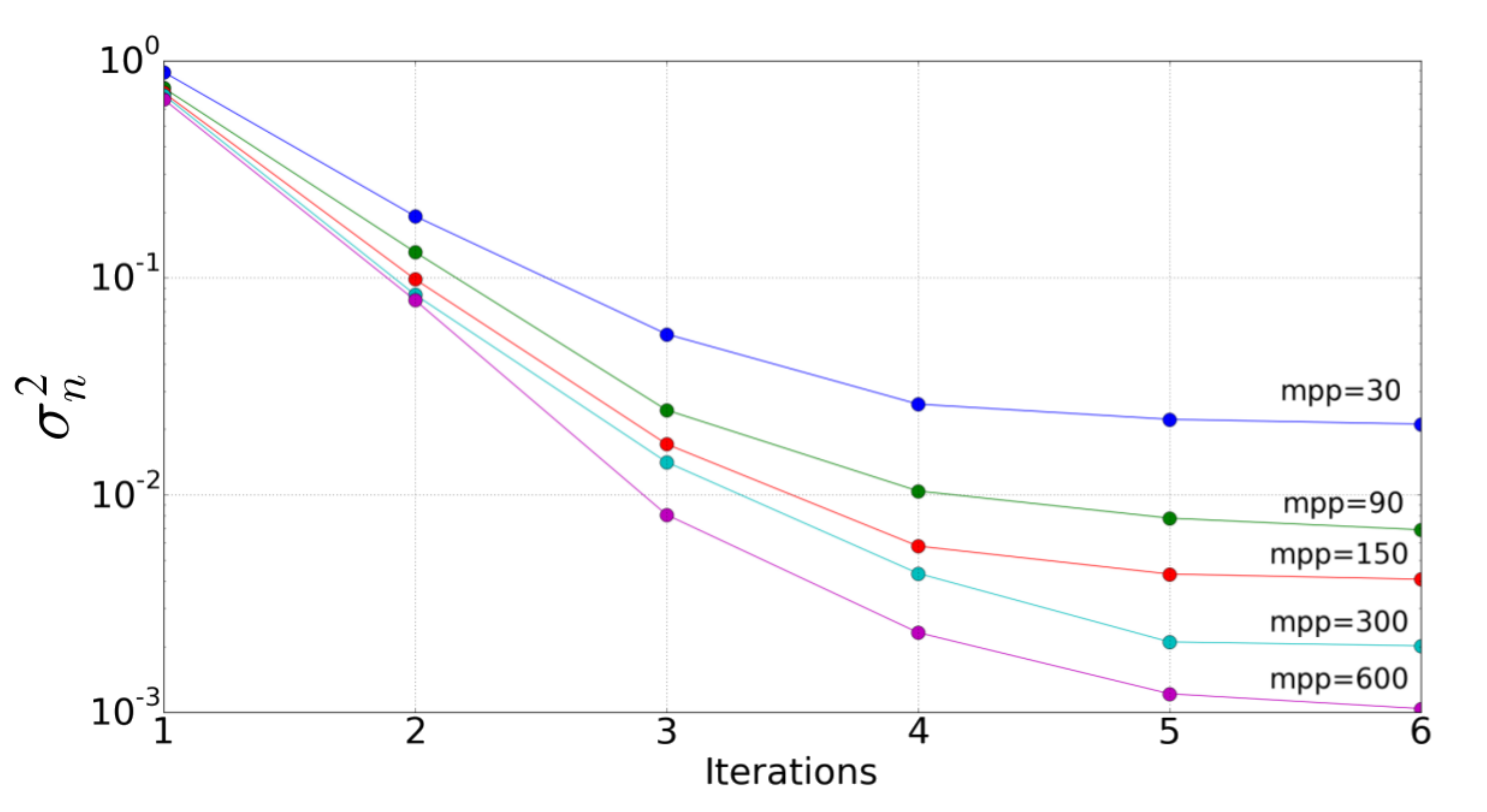}
		\captionsetup{width=0.95\linewidth, labelfont=bf}
		\caption{Behaviour of the variance for increased number of iterations of the PHOM algorithm. In each iteration the mean values of each stabiliser combination are reconstructed by performing a scan measuring an interval of length $\pi$ divided in ten points and performing a fixed amount of measurements per point (mpp) for each of these divisions. The variance approaches a constant value after several iterations. This is due to the finite number of mpp introducing a constant statistical noise. As the number of mpp increases this noise decreases.}
		\label{PHOM_fixedMPP_changeIterations}
	\end{center}
\end{figure}

\subsection{PHOM simulations}

In this section we discuss numerical simulations to obtain the behaviour of the variances for the PHOM with the number of measurements used in the three-plaquette case.

We initially fix the number of iterations $I$ and we choose a number $n$ of copies of the initial quantum state that we can measure. We then choose $\mathcal{N}$ different 7-dimensional vectors representing the seven initially unknown phases $\boldsymbol{\phi}^{(k)}$ ($k=1,...,\mathcal{N}$). We perform the PHOM following Sec.~\ref{PHOM} and the expressions \eqref{3plaqlikelihood} to \eqref{generalizemeans} in \ref{StabilizerAppendix} in order to obtain an estimate of the seven phases $\boldsymbol{\phi}^{(k)}_{\text{est}}$. For each of the vectors, we calculate the difference between the estimated phases and the phases chosen initially: $\boldsymbol{\phi}^{(k)}_{\text{est}}-\boldsymbol{\phi}^{(k)}$. We finally compute an estimate of the variance of the PHOM for a given $n$ and a fixed number of iterations $I$ as
\begin{equation}
\sigma^2_n=\frac{1}{\mathcal{N}}\sum_{k=1}^{\mathcal{N}}(\boldsymbol{\phi}^{(k)}_{\text{est}}-\boldsymbol{\phi}^{(k)})^2,
\end{equation}
where $\mathcal{N}$ is a large number to obtain a good estimate of $\sigma^2_n$, in our case $\mathcal{N}=50000$. To perform the scans of each stabiliser we divide the intervals $[-\pi,\pi]$ of the angles $\theta$s into $M = 10$ points. Since we have to measure seven stabilisers, the number of measurements per point (mpp) we can perform is $\text{mpp}=n/(7MI)$.

Repeating this process using a different $n$ and $I = 1,\dots,4$ yields the data plotted with circles  in Fig.~\ref{Variance_Comparison_3Plaquettes}. A similar process is done for the constant cosine PHOM without performing iterations since they are not needed for this method (see Sec.~\ref{PHOM}). The results obtained for this case are represented by the black circles in Fig.~\ref{Variance_Comparison_3Plaquettes}. It can be seen that as the number of iterations used for the PHOM increases, the variance with the number of resources decreases until reaching a similar behaviour as that of the constant cosine PHOM. The reason for this is that the PHOM has two limitations, one given by the number of iterations and another by the finite number of measurements used for each scan. If a small number of iterations is used, most of the PHOM simulations do not converge and the result is a wrong estimation for the phases which results in a big variance. As the number of iterations increases, more simulations converge to a correct phase and the variance obtained decreases. After performing enough iterations, all the simulations converge and the only source of error is the number of measurements used for each scan. As this number increases, the statistical error of each scan performed decreases yielding a better estimate of each phase. This behaviour is represented in Fig.~\ref{PHOM_fixedMPP_changeIterations}.

We perform a fit of the relation between the variance and the number of measurements for the constant cosine PHOM and the PHOM when enough iterations are performed for it to converge. This fit yields $\sigma^2_n\approx 224/n$ for these cases.

\subsection{Marginal likelihood Bayes simulations}

To obtain the behaviour of the variance with the number of measurements for the marginal likelihood Bayes inference method we apply a generalisation of the method shown in the pseudocode of Fig.~\ref{pseudocode} with the expressions \eqref{marginalbayes1} to \eqref{generalseven} in \ref{StabilizerAppendix}. After performing enough measurements ($\approx 100$) the probability distribution obtained can be approximated by a normal probability distribution for each of the phases that is used for computing the variances and for estimating the error of this adaptive technique. The results for the variances are shown in Fig.~\ref{Variance_Comparison_3Plaquettes} where the data is plotted with blue squares. It is possible to see that the variance decreases as the number of measurements used increases. A numerical fit of the data obtained reveals that the variance decreases as $\sigma^2_n\approx16/n$, as was expected from the analytical derivation of the method in Sec.~\ref{3plaquettesec}.

\section{Conclusions and outlook}

In this work, we have introduced an adaptive Bayesian method to measure systematic phase shift errors appearing in the experimental preparation of multi-qubit states, in particular, we have used it for correcting the errors appearing in the logical states of the Steane code. This method is capable of finding the experimental configuration that optimises the information gained by each measurement performed. An analytical development of this method that yields a simple rule for this adaptive selection has been shown, thus saving computational power that would be otherwise used in finding numerically the optimal measurement at each step of the process.

\begin{figure}
	\begin{center}
		
		\includegraphics[width=0.7\textwidth]{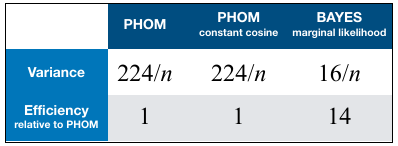}
		\captionsetup{width=0.95\linewidth, labelfont=bf}
		\caption{Comparison of the efficiencies among the different methods studied to measure phases for the three-plaquette case. For the PHOM case the results for the optimal selection of iterations have been used. }
		\label{table_efficiency}
	\end{center}
\end{figure}

We compared our method to the Phase Optimisation Method (PHOM), a non-adaptive phase estimation method based on a generalization of a Ramsey experiment for multi-qubit states that was recently realised for the implementation of the Steane code~\cite{nigg2014quantum}. We simulated both of them to measure quantum phases appearing in the preparation of quantum states needed for the Steane code. The efficiency obtained by simulation of our method is in agreement with the efficiency derived from the theoretical calculations, and shows an improvement by a reduction of the required measurement time by more than one order of magnitude when compared with the efficiency of the PHOM (see Fig.~\ref{table_efficiency}).

Furthermore, the method, illustrated for the optimisation of the seven-qubit Steane code, is applicable to other QEC codes and stabiliser states. Additionally, since this method only relies on the application of single-qubit rotations and measurements, it can correct systematic errors appearing in multi-qubit states implemented in different physical platforms. Thus, it has potential application in a variety of systems for quantum information processing such as, e.g., trapped ions, Rydberg atoms in optical lattices or tweezer arrays or other AMO or solid-state architectures.

\section*{Acknowledgments}
We thank P. Schindler for useful discussions. We acknowledge support by U.S. A.R.O. through Grant No. W911NF-14-1-010. The research is also based upon work supported by the Office of the Director of National Intelligence (ODNI), Intelligence Advanced Research Projects Activity (IARPA), via the U.S. Army Research Office Grant No. W911NF-16-1-0070. The views and conclusions contained herein are those of the authors and should not be interpreted as necessarily representing the official policies or endorsements, either expressed or implied, of the ODNI, IARPA, or the U.S. Government. The U.S. Government is authorized to reproduce and distribute reprints for Governmental purposes notwithstanding any copyright annotation thereon. Any opinions, findings, and conclusions or recommendations expressed in this material are those of the author(s) and do not necessarily reflect the view of the U.S. Army Research Office. We acknowledge the resources and support of High Performance Computing Wales, where most of the simulations were performed.

\appendix
\section{Three-plaquette case: Quantum states, likelihoods and pseudocode generalisation}
\label{StabilizerAppendix}
In this Appendix we provide details on the expressions appearing in the three-plaquette case. For the seven-qubit Steane code the logical 0 is given by
\begin{align}
\ket{0}_L=&\frac{1}{2\sqrt{2}} (\ket{0000000}+\ket{0110110}+\ket{1111000}+\ket{1001110}\nonumber\\ &\quad+\ket{0011011}+\ket{0101101}+\ket{1100011}+\ket{1010101})
\end{align}
Due to experimental errors, phases appear in the preparation of the $\ket{0}_L$ state. The state obtained will be then
\begin{align}
\ket{0'}_L=&\frac{1}{2\sqrt{2}} (\ket{0000000}+e^{i\phi_1}\ket{0110110}+e^{i\phi_2}\ket{1111000}+e^{i\phi_3}\ket{1001110}\nonumber\\ &+e^{i\phi_4}\ket{0011011}+e^{i\phi_5}\ket{0101101}+e^{i\phi_6}\ket{1100011}+e^{i\phi_7}\ket{1010101})
\end{align}
Single-qubit Z rotations can be performed on the seven qubits of the code state $\ket{0'}_L$ to obtain the following state and likelihoods for each stabiliser combination
	\begin{align}
	\label{3plaquettestate}
	&\ket{0'}_L\xrightarrow{\text{Rotations}} \frac{1}{2\sqrt{2}} (\ket{0000000}+e^{i[\phi_1+2(\theta_2+\theta_3+\theta_5+\theta_6)]}\ket{0110110}\nonumber\\&+e^{i[\phi_2+2(\theta_1+\theta_2+\theta_3+\theta_4)]}\ket{1111000}+e^{i[\phi_3+2(\theta_1+\theta_4+\theta_5+\theta_6)]}\ket{1001110}\nonumber\\&+e^{i[\phi_4+2(\theta_3+\theta_4+\theta_6+\theta_7)]}\ket{0011011}+e^{i[\phi_5+2(\theta_2+\theta_4+\theta_5+\theta_7)]}\ket{0101101}\nonumber\\&+e^{i[\phi_6+2(\theta_1+\theta_2+\theta_6+\theta_7)]}\ket{1100011}+e^{i[\phi_7+2(\theta_1+\theta_3+\theta_5+\theta_7)]}\ket{1010101}),
	\end{align}
\begingroup
\allowdisplaybreaks
	\begin{align}
	\label{3plaqlikelihood}
&P_1(\pm_{\boldsymbol{\theta}} |\boldsymbol{\phi})=\frac{1}{8}\{ 4\pm\cos[\phi_2+2(\theta_1+\theta_2+\theta_3+\theta_4)]\pm\cos[\phi_1-\phi_3+2(-\theta_1+\theta_2+\theta_3-\theta_4)]\nonumber\\&\pm\cos[\phi_4-\phi_6+2(-\theta_1-\theta_2+\theta_3+\theta_4)]\pm\cos[\phi_5-\phi_7+2(-\theta_1+\theta_2-\theta_3+\theta_4)]\},\\
	&P_2(\pm_{\boldsymbol{\theta}} |\boldsymbol{\phi})=\frac{1}{8}\{ 4\pm\cos[\phi_1+2(\theta_2+\theta_3+\theta_5+\theta_6)]\pm\cos[\phi_2-\phi_3+2(\theta_2+\theta_3-\theta_5-\theta_6)]\nonumber\\&\pm\cos[\phi_4-\phi_5+2(-\theta_2+\theta_3-\theta_5+\theta_6)]\pm\cos[\phi_6-\phi_7+2(\theta_2-\theta_3-\theta_5+\theta_6)]\},\\
	&P_3(\pm_{\boldsymbol{\theta}} |\boldsymbol{\phi})=\frac{1}{8}\{ 4\pm\cos[\phi_4+2(\theta_3+\theta_4+\theta_6+\theta_7)]\pm\cos[\phi_1-\phi_5+2(\theta_3-\theta_4+\theta_6-\theta_7)]\nonumber\\&\pm\cos[\phi_2-\phi_6+2(\theta_3+\theta_4-\theta_6-\theta_7)]\pm\cos[\phi_3-\phi_7+2(-\theta_3+\theta_4+\theta_6-\theta_7)]\},\\&P_{12}(\pm_{\boldsymbol{\theta}} |\boldsymbol{\phi})=\frac{1}{8}\{ 4\pm\cos[\phi_3+2(\theta_1+\theta_4+\theta_5+\theta_6)]\pm\cos[\phi_1-\phi_2+2(-\theta_1-\theta_4+\theta_5+\theta_6)]\nonumber\\&\pm\cos[\phi_4-\phi_7+2(-\theta_1+\theta_4-\theta_5+\theta_6)]\pm\cos[\phi_5-\phi_6+2(-\theta_1+\theta_4+\theta_5-\theta_6)],\\
	&P_{13}(\pm_{\boldsymbol{\theta}} |\boldsymbol{\phi})=\frac{1}{8}\{ 4\pm\cos[\phi_6+2(\theta_1+\theta_2+\theta_6+\theta_7)]\pm\cos[\phi_1-\phi_7+2(-\theta_1+\theta_2+\theta_6-\theta_7)]\nonumber\\&\pm\cos[\phi_2-\phi_4+2(\theta_1+\theta_2-\theta_6-\theta_7)]\pm\cos[\phi_3-\phi_5+2(\theta_1-\theta_2+\theta_6-\theta_7)]\},\\
	&P_{23}(\pm_{\boldsymbol{\theta}} |\boldsymbol{\phi})=\frac{1}{8}\{ 4\pm\cos[\phi_5+2(\theta_2+\theta_4+\theta_5+\theta_7)]\pm\cos[\phi_1-\phi_4+2(\theta_2-\theta_4+\theta_5-\theta_7)]\nonumber\\&\pm\cos[\phi_2-\phi_7+2(\theta_2+\theta_4-\theta_5-\theta_7)]\pm\cos[\phi_3-\phi_6+2(-\theta_2+\theta_4+\theta_5-\theta_7)]\},\\
	&P_{123}(\pm_{\boldsymbol{\theta}} |\boldsymbol{\phi})=\frac{1}{8}\{ 4\pm\cos[\phi_7+2(\theta_1+\theta_3+\theta_5+\theta_7)]\pm\cos[\phi_1-\phi_6+2(-\theta_1+\theta_3+\theta_5-\theta_7)]\nonumber\\&\pm\cos[\phi_2-\phi_5+2(\theta_1+\theta_3-\theta_5-\theta_7)]\pm\cos[\phi_3-\phi_4+2(\theta_1-\theta_3+\theta_5-\theta_7)]\}.
	\end{align}
	\endgroup
The expected values of each plaquette can be easily obtained from these expressions. For example, the first plaquette expected value is given by
\begin{equation}
\label{generalizemeans}
\langle S^{(1)}_x\rangle=P_1(+_{\boldsymbol{\theta}} |\boldsymbol{\phi})-P_1(-_{\boldsymbol{\theta}} |\boldsymbol{\phi})
\end{equation}
The expected values for the other plaquettes can be obtained from the corresponding likelihood in the same way. The marginal likelihoods are given by
\begin{align}
	\label{marginalbayes1}
	P_1(\pm_{\boldsymbol{\theta}} |\phi_2)=\frac{4\pm \cos[\phi_2-\tilde{\theta}_2]}{8}\\
	P_2(\pm_{\boldsymbol{\theta}} |\phi_1)=\frac{4\pm \cos[\phi_1-\tilde{\theta}_1]}{8}\\
	P_3(\pm_{\boldsymbol{\theta}} |\phi_4)=\frac{4\pm \cos[\phi_4-\tilde{\theta}_4]}{8}\\
	P_{12}(\pm_{\boldsymbol{\theta}} |\phi_3)=\frac{4\pm \cos[\phi_3-\tilde{\theta}_3]}{8}\\
	P_{13}(\pm_{\boldsymbol{\theta}} |\phi_6)=\frac{4\pm \cos[\phi_6-\tilde{\theta}_6]}{8}\\
	P_{23}(\pm_{\boldsymbol{\theta}} |\phi_5)=\frac{4\pm \cos[\phi_5-\tilde{\theta}_5]}{8}\\\label{marginalbayes2}
	P_{123}(\pm_{\boldsymbol{\theta}} |\phi_7)=\frac{4\pm \cos[\phi_7-\tilde{\theta}_7]}{8}
\end{align}
where
\begin{equation}
\tilde{\Theta}=\begin{cases}
\tilde{\theta}_2\equiv -2(\theta_1+\theta_2+\theta_3+\theta_4)\\
\tilde{\theta}_1\equiv -2(\theta_2+\theta_3+\theta_5+\theta_6)\\
\tilde{\theta}_4\equiv -2(\theta_3+\theta_4+\theta_6+\theta_7)\\
\tilde{\theta}_3\equiv -2(\theta_1+\theta_4+\theta_5+\theta_6)\\
\tilde{\theta}_6\equiv -2(\theta_1+\theta_2+\theta_6+\theta_7)\\
\tilde{\theta}_5\equiv -2(\theta_2+\theta_4+\theta_5+\theta_7)\\
\tilde{\theta}_7\equiv -2(\theta_1+\theta_3+\theta_5+\theta_7) \end{cases}
\end{equation}
As for the generalization of the pseudocode in Fig.~\ref{pseudocode}, it is achieved by changing the previous definitions to the following ones
\begin{align}
J&=1,...,7\nonumber,\\
S&=\set{\tilde{\theta}_i=\phi_i+\beta_i}\nonumber,\\
\mathcal{Q}&=\set{1,2,3,4,5,6,7}\nonumber,\\\label{generalseven}
\mathcal{T}&=\set{\theta_1,\theta_2,\theta_3,\theta_4,\theta_5,\theta_6,\theta_7},\\
\boldsymbol{\phi}&=\set{\phi_1,\phi_2,\phi_3,\phi_4,\phi_5,\phi_6,\phi_7}\nonumber,\\
\mathcal{P}&=\set{P(\phi_1),P(\phi_2),P(\phi_3),P(\phi_4),P(\phi_5),P(\phi_6),P(\phi_7)}\nonumber,\\
\mathcal{S}&=\{S^{(2)}_x,S^{(1)}_x,S^{(1)}_xS^{(2)}_x,S^{(3)}_x,S^{(2)}_xS^{(3)}_x,S^{(1)}_xS^{(3)}_x,S^{(1)}_xS^{(2)}_xS^{(3)}_x\}\nonumber,\\
\Pi&=\{P_2(\pm_{\boldsymbol{\theta}}|\phi_1),P_1(\pm_{\boldsymbol{\theta}}|\phi_2),P_{12}(\pm_{\boldsymbol{\theta}}|\phi_3),P_3(\pm_{\boldsymbol{\theta}}|\phi_4),P_{23}(\pm_{\boldsymbol{\theta}}|\phi_5),P_{13}(\pm_{\boldsymbol{\theta}}|\phi_6),P_{123}(\pm_{\boldsymbol{\theta}}|\phi_7)\}\nonumber,
\end{align}
where the $\beta_i$ in $S$ are chosen randomly from the set $\{+\pi/2, -\pi/2\}$.

\section{Analytical study of the scaling of the variance}
\label{VarianceBehaviour}

In this Appendix, we present some details on the scaling of the variance with the number of measurements when applying the Bayesian adaptive method to the single-qubit state of Eq.~\eqref{qubitstate}. The expressions obtained are easy to generalize to the two and three-plaquette cases. Let us suppose after $n$ measurements the knowledge about the phase $\phi$ is given by a normal distribution with mean value $\overline\phi_n$ and variance $\sigma^2_n$
\begin{equation}
\label{gaussianprob}
P_n(\phi)=\frac{1}{\sqrt{2\pi\sigma^2_n}}e^{-\frac{(\phi-\overline\phi_n)^2}{2\sigma^2_n}}.
\end{equation}
After performing a measurement the updated probability distribution is given by
\begin{equation}
P^\pm_{n+1}(\phi)=\frac{1}{p^\pm_n}\left( \frac{1\pm \cos(\phi-\theta_n)}{2}\right) P_n(\phi),
\end{equation}
where $p^\pm_n$ is the probability at step $n$ of obtaining a $+$ or $-$ in the measurement $n+1$. Since at step $n$ it is unknown what the measurement $n+1$ will yield, we consider the expected value of the variance after the measurement $n+1$, $\overline{\sigma^2}_{n+1}$
\begin{equation}
\label{sigmameanapp}
\overline{\sigma^2}_{n+1}=p^+_n(\sigma^+_{n+1})^2+p^-_n(\sigma^-_{n+1})^2,
\end{equation}
where $(\sigma^\pm_{n+1})^2$ are the variances after obtaining a $+$ or $-$ for the measurement $n+1$ given by
\begin{equation}
\label{sdpm}
({\sigma^\pm}_{n+1})^2=\int_{-\pi}^{\pi}\phi^2P^\pm_{n+1}(\phi)d\phi-\left(\int_{-\pi}^{\pi}\phi P^\pm_{n+1}(\phi)d\phi \right)^2.
\end{equation}
This yields
\begin{equation}
\label{positivesigma}
p^+_n(\sigma^+_{n+1})^2=\int_{-\pi}^{\pi}\phi^2P_n(\phi)\frac{1+\cos(\phi-\theta_n)}{2}d\phi-\frac{1}{p^+_n}\left(\int_{-\pi}^{\pi}\phi P_n(\phi)\frac{1+\cos(\phi-\theta_n)}{2}d\phi\right)^2,
\end{equation}
\begin{equation}
\label{negativesigma}
p^-_n(\sigma^-_{n+1})^2=\int_{-\pi}^{\pi}\phi^2P_n(\phi)\frac{1-\cos(\phi-\theta_n)}{2}d\phi-\frac{1}{p^+_n}\left(\int_{-\pi}^{\pi}\phi P_n(\phi)\frac{1-\cos(\phi-\theta_n)}{2}d\phi\right)^2.
\end{equation}
Introducing \eqref{positivesigma} and \eqref{negativesigma} into \eqref{sigmameanapp} yields
\begin{align}
\overline{\sigma^2}_{n+1}=&\ \sigma^2_n-\frac{1}{4p^+_np^-_n}\left(\int_{-\pi}^{\pi}\phi P_n(\phi)\cos(\phi-\theta_n)d\phi  \right)^2+\frac{p^+_n-p^-_n}{2p^+_np^-_n}\left(\int_{-\pi}^{\pi}\phi P_n(\phi)d\phi  \right)\nonumber\\&\times\left( \int_{-\pi}^{\pi}\phi P_n(\phi)\cos(\phi-\theta_n)d\phi\right)+\frac{4p^+_np^-_n-1}{4p^+_np^-_n}\left( \int_{-\pi}^{\pi}\phi P_n(\phi)d\phi\right)^2
\end{align}
where
\begin{align}
4p^+_np^-_n-1&=-\left( \int_{-\pi}^{\pi}P_n(\phi)\cos(\phi-\theta_n)d\phi\right)^2\\
p^+_n-p^-_n&=\left( \int_{-\pi}^{\pi}P_n(\phi)\cos(\phi-\theta_n)d\phi\right)
\end{align}
Thus, we obtain
\begin{align}
\label{sdstepbystepint}
\overline{{\sigma^2}_{n+1}-{\sigma^2}_{n}}=-\frac{1}{p^+p^-}\bigg[ \left( \int_{-\pi}^{\pi}\frac{P_{n}(\phi)\cos(\phi-\theta_n)}{2}d\phi \right) \nonumber\\\times\left( \int_{-\pi}^{\pi}\phi P_{n}(\phi)d\phi \right)-\left( \int_{-\pi}^{\pi}\frac{\phi P_{n}(\phi)\cos(\phi-\theta_n)}{2}d\phi \right) \bigg]^2
\end{align}
If we consider $P_n(\phi)$ has a low standard deviation we can change the intervals of integration $[-\pi,\pi)$ with $(-\infty,\infty)$ to obtain
\begin{align}
&\int_{-\infty}^{\infty}\phi P_{n}(\phi)d\phi=\overline{\phi}_n,\nonumber\\
&\int_{-\infty}^{\infty}\frac{P_{n}(\phi)\cos(\phi-\theta_n)}{2}d\phi=\frac{e^{-\sigma^2_n/2}}{2}\cos(\overline{\phi}_n-\theta_n),\nonumber\\
&\int_{-\infty}^{\infty}\frac{\phi P_{n}(\phi)\cos(\phi-\theta_n)}{2}d\phi=-\frac{e^{-\sigma^2_n/2}\sigma^2_n}{2}\sin(\overline{\phi}_n-\theta_n)+\frac{e^{-\sigma^2_n/2}\bar{\phi}_n}{2}\cos(\overline{\phi}_n-\theta_n),\nonumber\\
& p^\pm=\frac{1}{2}\left(1\pm e^{-\sigma^2_n/2}\cos(\overline{\phi}_n-\theta_n)\right).
\end{align}
Replacing the values of these integrals in \eqref{sdstepbystepint} we obtain
\begin{equation}
\overline{{\sigma^2}_{n+1}-{\sigma^2}_{n}}=-\frac{e^{-\sigma^2_n}\sigma^4_n\sin^2(\overline{\phi}_n-\theta_n)}{1-e^{-\sigma^2_n}\cos^2(\overline{\phi}_n-\theta_n)}.
\end{equation}
Similar calculations can be performed for the two and three plaquette case likelihoods, the only difference being the cosine appearing in the likelihood having amplitude $1/4$ and $1/8$ respectively and the angle $\tilde\theta_{i,n}$ being a linear combination of the rotations performed on different qubits at the step $n$. Taking this into account, for the two plaquette case the decrease in the variance for each $\phi_i$ is
\begin{equation}
\overline{{\sigma^2}_{i,n+1}-{\sigma^2}_{i,n}}=-\frac{e^{-\sigma^2_{i,n}}\sigma^4_{i,n}\sin^2(\overline{\phi}_{i,n}-\theta_{i,n})}{4-e^{-\sigma^2_{i,n}}\cos^2(\overline{\phi}_{i,n}-\theta_{i,n})}
\end{equation}
and for the three plaquette case
\begin{equation}
\overline{{\sigma^2}_{i,n+1}-{\sigma^2}_{i,n}}=-\frac{e^{-\sigma^2_{i,n}}\sigma^4_{i,n}\sin^2(\overline{\phi}_{i,n}-\theta_{i,n})}{16-e^{-\sigma^2_{i,n}}\cos^2(\overline{\phi}_{i,n}-\theta_{i,n})}.
\end{equation}

\section*{References}

\providecommand{\newblock}{}


\begin{thebibliography}{10}
	\expandafter\ifx\csname url\endcsname\relax
	\def\url#1{{\tt #1}}\fi
	\expandafter\ifx\csname urlprefix\endcsname\relax\def\urlprefix{URL }\fi
	\providecommand{\eprint}[2][]{\url{#2}}
	
	\bibitem{nielsen2002quantum}
	Nielsen M~A and Chuang I 2002 {\em Quantum Computation and Quantum
		Information\/} (Cambridge University Press)
	
	\bibitem{divincenzo2000physical}
	DiVincenzo D~P 2000 {\em Fortschr. Phys.: Progress of Physics\/} {\bf 48}
	771--783
	
	\bibitem{terhal2015quantum}
	Terhal B~M 2015 {\em Rev. Mod. Phys.\/} {\bf 87} 307
	
	\bibitem{breuer2002theory}
	Breuer H~P and Petruccione F 2002 {\em The theory of open quantum systems\/}
	(Oxford University Press)
	
	\bibitem{ramsey1950molecular}
	Ramsey N~F 1950 {\em Phys. Rev.\/} {\bf 78} 695
	
	\bibitem{foot2005atomic}
	Foot C~J 2005 {\em Atomic Physics\/} (Oxford University Press)
	
	\bibitem{huszar2012adaptive}
	Husz{\'a}r F and Houlsby N~M 2012 {\em Phys. Rev. A\/} {\bf 85} 052120
	
	\bibitem{wiebe2016efficient}
	Wiebe N and Granade C 2016 {\em Phys. Rev. Lett.\/} {\bf 117} 010503
	
	\bibitem{granade2017practical}
	Granade C, Ferrie C and Flammia S~T 2017 {\em New J. Phys.\/} {\bf 19} 113017
	
	\bibitem{sugiyama2012adaptive}
	Sugiyama T, Turner P~S and Murao M 2012 {\em Phys. Rev. A\/} {\bf 85} 052107
	
	\bibitem{fischer2000quantum}
	Fischer D~G, Kienle S~H and Freyberger M 2000 {\em Phys. Rev. A\/} {\bf 61}
	032306
	
	\bibitem{kravtsov2013experimental}
	Kravtsov K, Straupe S, Radchenko I, Houlsby N, Husz{\'a}r F and Kulik S 2013
	{\em Phys. Rev. A\/} {\bf 87} 062122
	
	\bibitem{mahler2013adaptive}
	Mahler D, Rozema L~A, Darabi A, Ferrie C, Blume-Kohout R and Steinberg A 2013
	{\em Physical review letters\/} {\bf 111} 183601
	
	\bibitem{granade2012robust}
	Granade C~E, Ferrie C, Wiebe N and Cory D~G 2012 {\em New Journal of Physics\/}
	{\bf 14} 103013
	
	\bibitem{lidar2013quantum}
	Lidar D~A and Brun T~A 2013 {\em Quantum error correction\/} (Cambridge
	University Press)
	
	\bibitem{cramer2010efficient}
	Cramer M, Plenio M~B, Flammia S~T, Somma R, Gross D, Bartlett S~D,
	Landon-Cardinal O, Poulin D and Liu Y~K 2010 {\em Nat. Commun.\/} {\bf 1} 149
	
	\bibitem{muller2016iterative}
	M{\"u}ller M, Rivas A, Mart{\'\i}nez E, Nigg D, Schindler P, Monz T, Blatt R
	and Martin-Delgado M 2016 {\em Phys. Rev. X\/} {\bf 6} 031030
	
	\bibitem{steane1996error}
	Steane A~M 1996 {\em Phys. Rev. Lett.\/} {\bf 77} 793
	
	\bibitem{haffner2008quantum}
	H{\"a}ffner H, Roos C~F and Blatt R 2008 {\em Phys. Rep.\/} {\bf 469} 155--203
	
	\bibitem{nigg2014quantum}
	Nigg D, M{\"u}ller M, Martinez E~A, Schindler P, Hennrich M, Monz T,
	Martin-Delgado M~A and Blatt R 2014 {\em Science\/}  1253742
	
	\bibitem{sriarunothai2018speeding}
	Sriarunothai T, W{\"o}lk S, Giri G~S, Friis N, Dunjko V, Briegel H~J and
	Wunderlich C 2018 {\em Quantum Science and Technology\/} {\bf 4} 015014
	
	\bibitem{johanning2009quantum}
	Johanning M, Var{\'o}n A~F and Wunderlich C 2009 {\em Journal of Physics B:
		Atomic, Molecular and Optical Physics\/} {\bf 42} 154009
	
	\bibitem{jaksch2000fast}
	Jaksch D, Cirac J, Zoller P, Rolston S, C{\^o}t{\'e} R and Lukin M 2000 {\em
		Phys. Rev. Lett.\/} {\bf 85} 2208
	
	\bibitem{saffman2010quantum}
	Saffman M, Walker T~G and M{\o}lmer K 2010 {\em Rev. Mod. Phys.\/} {\bf 82}
	2313
	
	\bibitem{crow2016numerical}
	Crow D, Joynt R and Saffman M 2016 {\em Phys. Rev. Lett.\/} {\bf 117}(13)
	130503
	
	\bibitem{anderson2011trapping}
	Anderson S~E, Younge K and Raithel G 2011 {\em Phys. Rev. Lett.\/} {\bf 107}
	263001
	
	\bibitem{viteau2011rydberg}
	Viteau M, Bason M, Radogostowicz J, Malossi N, Ciampini D, Morsch O and
	Arimondo E 2011 {\em Phys. Rev. Lett.\/} {\bf 107} 060402
	
	\bibitem{schauss2012observation}
	Schau{\ss} P, Cheneau M, Endres M, Fukuhara T, Hild S, Omran A, Pohl T, Gross
	C, Kuhr S and Bloch I 2012 {\em Nature\/} {\bf 491} 87
	
	\bibitem{nogrette2014single}
	Nogrette F, Labuhn H, Ravets S, Barredo D, B{\'e}guin L, Vernier A, Lahaye T
	and Browaeys A 2014 {\em Phys. Rev. X\/} {\bf 4} 021034
	
	\bibitem{xia2015randomized}
	Xia T, Lichtman M, Maller K, Carr A, Piotrowicz M, Isenhower L and Saffman M
	2015 {\em Phys. Rev. Lett.\/} {\bf 114} 100503
	
	\bibitem{hanson2008coherent}
	Hanson R and Awschalom D~D 2008 {\em Nature\/} {\bf 453} 1043
	
	\bibitem{corcoles2015demonstration}
	C{\'o}rcoles A~D, Magesan E, Srinivasan S~J, Cross A~W, Steffen M, Gambetta J~M
	and Chow J~M 2015 {\em Nat. Commun.\/} {\bf 6} 6979
	
	\bibitem{gambetta2017building}
	Gambetta J~M, Chow J~M and Steffen M 2017 {\em npj Quantum Inf.\/} {\bf 3} 2
	
	\bibitem{kelly2015state}
	Kelly J, Barends R, Fowler A~G, Megrant A, Jeffrey E, White T~C, Sank D, Mutus
	J~Y, Campbell B, Chen Y {\em et~al.\/} 2015 {\em Nature\/} {\bf 519} 66
	
	\bibitem{waldherr2014quantum}
	Waldherr G, Wang Y, Zaiser S, Jamali M, Schulte-Herbr{\"u}ggen T, Abe H,
	Ohshima T, Isoya J, Du J, Neumann P {\em et~al.\/} 2014 {\em Nature\/} {\bf
		506} 204
	
	\bibitem{fedorov2012implementation}
	Fedorov A, Steffen L, Baur M, da~Silva M~P and Wallraff A 2012 {\em Nature\/}
	{\bf 481} 170
	
	\bibitem{negnevitsky2018feedback}
	Negnevitsky V 2018 {\em Feedback-stabilised quantum states in a mixed-species
		ion system\/} Ph.D. thesis ETH Zurich
	
	\bibitem{higgins2007entanglement}
	Higgins B~L, Berry D~W, Bartlett S~D, Wiseman H~M and Pryde G~J 2007 {\em
		Nature\/} {\bf 450} 393
	
	\bibitem{giovannetti2004quantum}
	Giovannetti V, Lloyd S and Maccone L 2004 {\em Science\/} {\bf 306} 1330--1336
	
	\bibitem{paesani2017experimental}
	Paesani S, Gentile A~A, Santagati R, Wang J, Wiebe N, Tew D~P, O'Brien J~L and
	Thompson M~G 2017 {\em Phys. Rev. Lett.\/} {\bf 118} 100503
	
	\bibitem{bombin2006topological}
	Bombin H and Martin-Delgado M~A 2006 {\em Phys. Rev. Lett.\/} {\bf 97}(18)
	180501
	
\end{thebibliography}
\end{document}